\documentclass[review]{elsarticle}

\journal{}

\usepackage{enumitem}
\usepackage{colortbl}
\usepackage{array}
\usepackage{subfig}
\usepackage{braket}
\usepackage{amsmath}
\usepackage{amsfonts}
\usepackage{amssymb}
\usepackage{graphicx}
\usepackage{qcircuit}
\usepackage{tikz}
\usetikzlibrary{tikzmark}
\usepackage{dcolumn}
\usepackage{tikz}
\usepackage{mathtools}
\usetikzlibrary{arrows.meta}
\usepackage{float}
\usetikzlibrary{positioning,shapes,arrows}
\usepackage{soul}
\DeclarePairedDelimiter{\ceil}{\lceil}{\rceil}
\makeatletter 
\newcount\SOUL@minus
\makeatother  

\begin{document}
\begin{frontmatter}

\title{Quantum circuit representation of Bayesian networks}

\author[address1]{Sima E. Borujeni}
\author[address1]{Saideep Nannapaneni\corref{mycorrespondingauthor}}
\cortext[mycorrespondingauthor]{Corresponding author. Phone: +1 316-978-6240}

\author[address2]{Nam H. Nguyen
\fnref{fn1}}
\fntext[fn1]{Present address: Boeing Research \& Technology, Huntington Beach, CA 92647, USA}
\author[address2]{Elizabeth C. Behrman}
\author[address3]{James E. Steck}

\address[address1]{Department of Industrial, Systems, and Manufacturing Engineering, Wichita State University\\
1845 Fairmount St, Box 35, Wichita, KS, 67260 USA\\
\{sxborujeni@shockers.wichita.edu, saideep.nannapaneni@wichita.edu\}}

\address[address2]{Department of Mathematics, Statistics, and Physics, Wichita State University\\
1845 Fairmount St, Box 33, Wichita, KS, 67260 USA\\
\{nhnguyen5@shockers.wichita.edu, elizabeth.behrman@wichita.edu\}\\}
\address[address3]{Department of Aerospace Engineering, Wichita State University\\
1845 Fairmount St, Box 44, Wichita, KS, 67260 USA\\
\{james.steck@wichita.edu\}}

\begin{abstract}
Probabilistic graphical models such as Bayesian networks are widely used to model stochastic systems to perform various types of analysis such as probabilistic prediction, risk analysis, and system health monitoring, which can become computationally expensive in large-scale systems. While demonstrations of true quantum supremacy remain rare, quantum computing applications managing to exploit the advantages of amplitude amplification have shown significant computational benefits when compared against their classical counterparts. 
We develop a systematic method for designing a quantum circuit to represent a generic discrete Bayesian network with nodes that may have two or more states, where nodes with more than two states are mapped to multiple qubits. The  marginal probabilities associated with root nodes (nodes without any parent nodes) are represented using rotation gates, and the conditional probability tables associated with non-root nodes are represented using controlled rotation gates. The controlled rotation gates with more than one control qubit are represented using ancilla qubits. The proposed approach is demonstrated for three examples: a 4-node oil company stock prediction, a 10-node network for liquidity risk assessment, and a 9-node naive Bayes classifier for bankruptcy prediction. The circuits were designed and simulated using Qiskit, a quantum computing platform that enables simulations and also has the capability to run on real quantum hardware. The results were validated against those obtained from classical Bayesian network implementations.
\end{abstract}


\end{frontmatter}
\section{Introduction}
\label{sec:introduction}

Bayesian Networks, also known as Bayesian belief networks, are probabilistic graphical models used to represent knowledge about an uncertain domain. A Bayesian network is represented as a directed acyclic graph with nodes and edges, where nodes represent the random variables and edges represent the probabilistic dependence between nodes \citep{murphy2002dynamic}. 

Bayesian networks are used to perform two types of analysis: forward analysis, which provides a probabilistic prediction of the lower-level nodes in the Bayesian networks using probability distributions of the higher-level nodes, and inverse analysis, which infers the values of higher-level nodes using data on the lower-level nodes. The inverse analysis is commonly referred to as Bayesian inference since the inference analysis is carried out using the Bayes theorem. The forward analysis is typically performed through Monte Carlo analysis and has been used to perform uncertainty propagation \citep{nannapaneni2016performance}, performance evaluation \citep{zhu2003application}, reliability and risk analysis \citep{garvey2015analytical}, and prognostics \citep{ferreiro2012application}, whereas the inverse analysis has been used for system identification \citep{lee2016bayesian}, health monitoring \citep{kothamasu2006system}, and system diagnostics \citep{li2017dynamic}. Bayesian networks have been used to carry out a variety of analyses in several domains of science and engineering such as mechanical \citep{xu2012intelligent}, aerospace \citep{li2017dynamic, nannapaneni2018automated}, and manufacturing systems \citep{buyukozkan2015assessment, nannapaneni2018real}, industrial systems \citep{cai2016real}, healthcare \citep{kahn1997construction, kalet2015bayesian}, infrastructure systems \citep{hosseini2016modeling, nannapaneni2017performance}, biomedical systems \citep{miyauchi2018bayesian}, and transportation \citep{sun2006bayesian, pettet2017incident}. 

Some of the issues with the current implementations of Bayesian networks is the high computational expense in the presence of large number of nodes (random variables) for both forward and inverse analyses.  \textcolor{black}{One possible way to obviate this difficulty might be to use the principles of quantum computing (sometimes referred to as quantum-assisted computing). This is because quantum computers make use of ``superposition" which is} \textcolor{black}{the ability of quantum systems to be simultaneously in multiple different states}. 

Several algorithms have been developed using the principles of quantum mechanics that have demonstrated superior computational performance over the corresponding classical algorithms, and the most notable of these are Shor's algorithm \citep{shor1994algorithms} and Grover's algorithm \citep{grover1996fast}. Shor's algorithm is used for the factorization of integers; this algorithm has exponential speedup when compared to the best known classical algorithms. Grover's algorithm is used for search in an unstructured search space (such as an unstructured database), and has quadratic speedup. Due to its computational benefits, the Grover's algorithm has been used as a sub-routine in the development of many quantum algorithms for classification such as quantum support vector machines \citep{rebentrost2014quantum}, for clustering such as quantum k-means clustering \citep{aimeur2007quantum}, and for combinatorial optimization \citep{baritompa2005grover}. With regard to Bayesian networks, Ozols et al \citep{ozols2013quantum} used the principles of amplitude amplification \citep{brassard2002quantum}  to develop an algorithm for Bayesian inference (inverse analysis) known as quantum rejection sampling, which is a quantum version of the rejection sampling algorithm \citep{gilks1992adaptive} used for inference in classical Bayesian networks. Woerner and Egger \citep{woerner2019quantum} used the principles of amplitude amplification and estimation \citep{brassard2002quantum} to perform risk analysis (forward analysis), and demonstrated it with two toy problems from the financial industry. In this paper, we consider the representation of Bayesian networks in a quantum computing paradigm to facilitate the use of those quantum algorithms for forward and inverse analyses.

There are primarily two types of architectures that have widely been used to realize quantum computing: quantum gate models \citep{dallaire2016quantum} and quantum annealing \citep{bunyk2014architectural}. The quantum gate architecture uses a series of quantum gates that act on individual qubits to achieve the desired computation. 

More details regarding qubits and gates are available in Sections \ref{subsec:qubit} and \ref{subsec:gates}. A quantum circuit is a graphical representation of the sequence of gates implemented on various qubits to do the desired computation. Quantum annealing architecture uses the principles of quantum annealing \citep{boixo2013experimental}, which is a quantum equivalent to classical simulated annealing algorithm, to make the desired computation. According to Ajagekar et al \citep{ajagekar2020quantum}, quantum annealing architecture is better suited for optimization problems whereas quantum gate architecture facilitates universal quantum computation. As the goal of this paper is the representation of Bayesian networks, we use the quantum gate architecture as opposed to the quantum annealing architecture.

Quantum Bayesian networks were first introduced by \mbox{\cite{tucci1995quantum}} as an analog to classical Bayesian networks . He proposed that the conditional probabilities in a classical Bayesian networks can be represented using quantum complex amplitudes. Tucci argued that there could be infinite possible quantum Bayesian network for a given classical Bayesian network; however, methods to realize a quantum Bayesian network for a classical Bayesian network were not available. \mbox{\cite{moreira2016quantum}} proposed quantum-like Bayesian networks, where the marginal and conditional probabilities were represented using quantum probability amplitudes. To determine the parameters of a quantum Bayesian network, a similarity heuristic method was used which considered similarity between two dimensional vectors corresponding to the two states of the random variables.

Previous work considered quantum Bayesian networks with binary variables only and considered heuristics approaches for their representation; however, this paper also considers representation of variables with more than two states. The approach presented in this paper does not consider any heuristics and can be used to represent any generic discrete quantum Bayesian network.

\cite{low2014quantum} discussed the principles of quantum circuit design to represent a Bayesian network with discrete nodes that have two states, and also discussed the circuit design for implementing quantum rejection sampling for inference. In this paper, we consider the representation of generic discrete Bayesian networks with nodes that may have two or more states, and also discuss the decomposition of complex gates using elementary gates (discussed in Section \ref{subsec:gates}) such that they can be implemented on available quantum computing platforms. 


\textbf{Paper Contributions:} The overall contributions made through this paper are:  (\textbf{1}) Decomposition of a multi-qubit gate into elementary gates to represent a discrete variable with more than two states; (\textbf{2}) A systematic procedure to design a quantum circuit to represent a generic discrete Bayesian network with nodes that may have two or more states; and (\textbf{3}) Illustration of the proposed quantum circuit representation to three Bayesian networks used for oil price stock prediction, liquidity risk assessment, and bankruptcy prediction, and validating the results against classical Bayesian network implementations.

\textbf{Paper Organization:} Section \ref{sec:background} provides a brief background to qubits, quantum gates, and Bayesian networks. Section \ref{sec:method} details the proposed methods for designing a quantum circuit to represent a Bayesian network. Section \ref{sec:examples} details the application of the proposed methods to three Bayesian networks from the financial industry followed by concluding remarks in Section \ref{sec:conclusion}.


\section{Background}
\label{sec:background}
In this section, we provide a brief background to qubits, different quantum gates that are used to perform qubit transformations, and Bayesian Networks.

\subsection{Qubit}
\label{subsec:qubit}

A qubit (or a quantum bit) is an elementary unit of information in quantum computing, similar to a classical bit (or simply a bit) in classical computing. A bit is always in one of the either two basis states, 0 and 1, whereas a qubit can be in both the basis states simultaneously. This property of a qubit is known as quantum superposition \mbox{\citep{nielsen2002quantum}}. In quantum computing, the Dirac notation is used to represent the two basis states as $\Ket{0}$ and $\Ket{1}$. In general, any two orthonormal states can be used as the basis states but the commonly used basis states or computational basis are $\Ket{0}$ and $\Ket{1}$. Eq. (\ref{eqn:basis}) provides their vector representations as

\begin{equation}
\label{eqn:basis}
\Ket{0} = \begin{bmatrix}
                 {1} \\
                 {0} \\
         \end{bmatrix} \hspace{2cm}
\Ket{1}= \begin{bmatrix}

                 {0} \\
                 {1} \\
        \end{bmatrix} 
\end{equation}

A general pure state of a qubit is a superposition, which is linear combination of the two basis states written as  $ \Ket{\Psi} = c_1 \Ket{0} + c_2 \Ket{1}$ or

\begin{equation}
        \begin{bmatrix}
            {c_1}\\
            {c_2}\\
            \end{bmatrix} = c_1\begin{bmatrix}
             {1} \\
             {0} \\
            \end{bmatrix}
        +c_2\begin{bmatrix}
             {0} \\
             {1} \\
            \end{bmatrix}
\end{equation}

\noindent where $c_1$ and $c_2$ are  complex numbers, which represent the probability amplitudes corresponding to $\Ket{0}$ and $\Ket{1}$ respectively. When a measurement is made, the qubit collapses to one of the two basis states. The probabilities of observing the qubit in $\Ket{0}$ and $\Ket{1}$ states are computed as the inner product of their probability amplitudes and their complex conjugates, represented as $c_1^{\dagger} c_1 = |c_1|^2$ and $c_2^{\dagger} c_2 = |c_2|^2$ respectively, where $c_1^{\dagger}$ and $c_2^{\dagger}$ are the complex conjugates of $c_1$ and $c_2$ respectively. Since a qubit can be either in $\Ket{0}$ or in $\Ket{1}$, the sum of their probabilities is equal to unity ($|c_1|^2  + |c_2|^2=1$) \mbox{\citep{kopczyk2018quantum}}. 

When multiple qubits are used in computing, their joint state can be obtained through a tensor product of individual qubits. If $\Ket{\Psi_1}=a_1\Ket{0}+a_2\Ket{1}$ and $\ket{\Psi_2} = b_1\Ket{0}+b_2\Ket{1}$ represent two qubits with real-valued probability amplitudes, then their joint state is represented as $\Ket{\Psi_1} \otimes \Ket{\Psi_2} = a_1b_1\Ket{00}+a_1b_2\Ket{01}+a_2b_1\Ket{10}+a_2b_2\Ket{11}$. But the joint state need not always be a product state (tensor product of individual states). There are states of the joint system that can not be written as a product of individual states; these are called entangled states \mbox{\citep{horodecki2009quantum}}. Entanglement is quantum correlation stronger than any possible classical correlation \mbox{\citep{luo2003correlation}}.

An example of an entangled state of two qubits is the Bell state, defined as $\dfrac{1}{\sqrt{2}} \Ket{00} + \dfrac{1}{\sqrt{2}} \Ket{11}$ \citep{nielsen2002quantum}. Clearly, it can not be written as a tensor product of two qubits, $\Ket{\Psi_1}=a_1\Ket{0}+a_2\Ket{1}$ and $\ket{\Psi_2} = b_1\Ket{0}+b_2\Ket{1}$, because then $a_1b_1 = a_2b_2 = \dfrac{1}{\sqrt{2}}$ but also $a_1b_2 = a_2b_1 = 0$. Thus, neither qubit has an individual state, but the two are perfectly correlated. If one of the two qubits is measured to be in $\Ket{0}$ state, then the other qubit is also in $\Ket{0}$ due to the presence of entanglement between them.

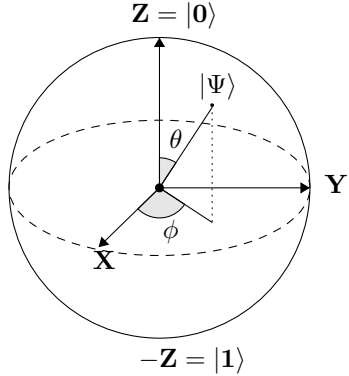
\begin{figure}
\begin{center}
\begin{tikzpicture}[line cap=round, line join=round, >=Triangle]
  \clip(-2.5,-2.49) rectangle (2.70,2.70);
  \draw [shift={(0,0)}, fill, fill, fill opacity=0.1] (0,0) -- (56.7:0.4) arc (56.7:90.:0.4) -- cycle;
  \draw [shift={(0,0)}, fill, fill, fill opacity=0.1] (0,0) -- (-135.7:0.4) arc (-135.7:-33.2:0.4) -- cycle;
  \draw(0,0) circle (2cm);
  \draw [rotate around={0.:(0.,0.)},dash pattern=on 3pt off 3pt] (0,0) ellipse (2cm and 0.9cm);
  \draw (0,0)-- (0.70,1.07);
  \draw [->] (0,0) -- (0,2);
  \draw [->] (0,0) -- (-0.81,-0.79);
  \draw [->] (0,0) -- (2,0);
  \draw [dotted] (0.7,1)-- (0.7,-0.46);
  \draw  (0,0)-- (0.7,-0.46);
  \draw (-0.08,-0.3) node[anchor=north west] {$\phi$};
  \draw (0.01,0.9) node[anchor=north west] {$\theta$};
  \draw (-1.01,-0.72) node[anchor=north west] {$\mathbf {X}$};
  \draw (2.07,0.3) node[anchor=north west] {$\mathbf {Y}$};
  \draw (-0.5,2.6) node[anchor=north west] {$\mathbf {Z=|0\rangle}$};
  \draw (-0.4,-2) node[anchor=north west] {$-\mathbf {Z=|1\rangle}$};
  \draw (0.4,1.65) node[anchor=north west] {$|\Psi\rangle$};
  \scriptsize
  \draw [fill] (0,0) circle (1.5pt);
  \draw [fill] (0.7,1.1) circle (0.5pt);
\end{tikzpicture}

\end{center}
\caption{Bloch sphere representation of a qubit}
\label{fig:Bloch}
\end{figure}

A Bloch sphere (shown in Figure \ref{fig:Bloch}) is often used for geometrical representation of a qubit \mbox{\citep{goyal2016geometry}}. In a Bloch sphere, the positive Z-axis corresponds to the $\Ket{0}$ state while the negative Z-axis corresponds to $\Ket{1}$ state. A pure state of a qubit is represented by a point on the Bloch sphere and can be represented as $\cos(\frac{\theta}{2}) \Ket{0} + e^{i\phi}\sin(\frac{\theta}{2}) \Ket{1}$ for given values of $(\theta,\phi)$. 

\subsection{Quantum gates}
\label{subsec:gates}

Quantum gates are mathematical operations performed on the qubits to change their probability amplitudes to gain the desired computations. The quantum gates are similar to classical gates (such as the AND gate) acting on classical bits. Geometrically, one-qubit gates represent unitary rotations about various axes in the Bloch sphere.

There are two elementary gates in quantum computing - $U_3$ and CNOT \citep{mckay2018qiskit}, which act on a single qubit and two qubits respectively. Any other multi-qubit gates can be decomposed into these elementary gates. We discuss in detail the one-qubit and two-qubit elementary gates. A more comprehensive review of gates is available in  \cite{nielsen2002quantum}.

 \subsubsection{One-qubit gates} We review the generic $U_3$ gate, and some of its special cases - $X$ (sometimes referred to as Pauli-$X$), $R_Y$ and $R_Z$ gates. $U_3$ gate has three parameters $\theta$, $\phi$ and $\lambda$, and it can be used to construct any arbitrary single qubit gate. The matrix representation of this gate is given as

\begin{equation}
\label{eqn:U_3matrix}
 U_3(\theta, \phi, \lambda) = \begin{bmatrix}
\cos\Big(\dfrac{\theta}{2}\Big) &  -e^{i\lambda} \sin\Big(\dfrac{\theta}{2}\Big) \\[2mm]
e^{i\phi} \sin\Big(\dfrac{\theta}{2}\Big) &  e^{i(\phi + \lambda)} \cos\Big(\dfrac{\theta}{2}\Big) \\
\end{bmatrix} 
\end{equation}

\noindent where $\theta$ represents the angle of rotation about the Y-axis, and $\phi$ and $\lambda$ represent the angles of rotation around the Z-axis. The generic $U_3$ gate is often represented as the $U$ gate.

\textbf{$R_Y$ gate:} The $R_Y$ gate is a single-qubit gate, which corresponds to a rotation of angle $\theta$ (radians) about the y-axis on the Bloch sphere. $R_Y$ gate can be represented as a special case of $U_3$ gate as
\begin{equation}
R_Y(\theta)=U_3(\theta, 0, 0)=
 \begin{bmatrix}
\cos\Big(\dfrac{\theta}{2}\Big) &  -\sin\Big(\dfrac{\theta}{2}\Big) \\[2mm]
\sin\Big(\dfrac{\theta}{2}\Big) & \cos\Big(\dfrac{\theta}{2}\Big)\\
\end{bmatrix}  
\end{equation}

\textbf{$R_Z$ gate} or \textbf{Phase-shift gate:}
$R_Z(\lambda)$ is another single qubit gate, which corresponds to rotation about the Z-axis by an angle $\lambda$ on the Bloch sphere. $R_Z$ gate can be represented as a special case of $U_3$ as

\begin{equation}
R_Z(\lambda)=U_3(0,0,\lambda)=
\begin{bmatrix}
1 &  \hspace{2mm} 0 \\
0 & \hspace{2mm} e^{i\lambda}\\
\end{bmatrix}   
\end{equation}

Using the matrix representations of $R_Y$ and $R_Z$ gates, the $U_3$ gate given in Eq. (\ref{eqn:U_3matrix}) can be decomposed into two phase-shift rotations and one rotation about the Y-axis as

\begin{equation}
\label{eq:U_3_3matricies}
\begin{aligned}
U_3(\theta, \phi, \lambda) &= \begin{bmatrix}
1 &  0 \\[2mm]
0 &  e^{i\phi} \\
\end{bmatrix} 
\begin{bmatrix}
\cos\Big(\dfrac{\theta}{2}\Big) &  -\sin\Big(\dfrac{\theta}{2}\Big) \\[2mm]
\sin\Big(\dfrac{\theta}{2}\Big) &  \cos\Big(\dfrac{\theta}{2}\Big) \\
\end{bmatrix} 
\begin{bmatrix}
1 &  0 \\[2mm]
0 &  e^{i\lambda} \\
\end{bmatrix} \\
& = R_Z(\phi)R_Y(\theta)R_Z(\lambda)
\end{aligned}
\end{equation}

For more detail about the rotation gates and decomposition of an arbitrary single qubit gate, refer to \cite{nielsen2002quantum} and \cite{cross2017open}.

\textbf{X gate}: The $X$ gate is the quantum-equivalent to the classical NOT gate or sometimes referred to as \textit{flip} gate, as it flips $\Ket{0}$ to $\Ket{1}$ and $\Ket{1}$ to $\Ket{0}$. The matrix notation of the $X$ gate is equal to 

\begin{equation}
 X = U_3\Big(\pi, -\frac{\pi}{2}, \frac{\pi}{2}\Big) = \begin{bmatrix}
0 &  \hspace{2mm} 1 \\
1 & \hspace{2mm} 0\\
\end{bmatrix}
\end{equation}

\subsubsection{Two-qubit gates} An elementary two-qubit gate is the controlled-NOT (CNOT or $CX$) gate. The two qubits on which the CNOT gate is implemented are referred to as the control qubit and target qubit. When the control qubit is $\Ket{0}$, the target qubit remains unchanged whereas when the control qubit is $\Ket{1}$, the $X$ gate is implemented on the target qubit. The CNOT gate does not have any effect on the control qubit \mbox{\citep{liu2008analytic}}. In the usual computational basis ($\Ket{00}, \Ket{01}, \Ket{10}, and \Ket{11}$ states),the matrix representation of a CNOT gate is given as 

\begin{equation}
    \centering
    CNOT=
    \begin{bmatrix}
    1 & 0 & 0 & 0 \\
    {0} & {1} & {0} & {0} \\
    {0} & {0} & {0} & {1} \\
    {0} & {0} & {1} & {0} \\
    \end{bmatrix}
\end{equation}

For example, consider two qubits given as $\Ket{\Psi_1}=a_1\Ket{0}+a_2\Ket{1}$ and $\ket{\Psi_2} = b_1\Ket{0}+b_2\Ket{1}$ on which a CNOT gate is implemented with $\Ket{\Psi_1}$ as the control qubit. The combined quantum state before the application of CNOT gate is given by their tensor product, $\Ket{\Psi_1} \otimes \Ket{\Psi_2} = a_1b_1 \Ket{00} + a_1b_2 \Ket{01} + a_2b_1 \Ket{10} + a_2b_2 \Ket{11}$. The quantum state after the application of the CNOT gate is equal to $a_1b_1 \Ket{00} + a_1b_2 \Ket{01} + a_2b_1 \Ket{11} + a_2b_2 \Ket{10}$. The $\Ket{00}$ and $\Ket{01}$ remain unchanged  since the control qubit is $\Ket{0}$ whereas $\Ket{10}$ and $\Ket{11}$ become $\Ket{11}$ and $\Ket{10}$ respectively as the $X$ gate is applied when the control qubit is $\Ket{1}$. Similar to the CNOT gate, we have the $CU$ gate, which implements the $U$ (or the $U_3$) gate when the control qubit is $\Ket{1}$. Given the matrix of the $U$ gate in Eq. (\ref{eqn:U_3matrix}), the matrix representation of $CU$ can be written as 

\begin{equation}
    CU=
    \begin{bmatrix}
    {1} & \hspace{4mm}{0} & {0} & {0} \\
    {0} & \hspace{4mm}{1} & {0} & {0} \\
    {0} & \hspace{4mm}{0} & \cos\Big(\dfrac{\theta}{2}\Big)& -e^{i\lambda} \sin\Big(\dfrac{\theta}{2}\Big) \\[2mm]
    {0} & \hspace{4mm} {0} & e^{i\phi} \sin\Big(\dfrac{\theta}{2}\Big)& e^{i(\phi + \lambda)} \cos\Big(\dfrac{\theta}{2}\Big) \\
    \end{bmatrix}
\end{equation}

A variant of the $CU$ gate is the $CR_Y(\theta)$ gate, which implements the rotation gate $R_Y(\theta)$ on the target qubit when the control qubit is equal to $\Ket{1}$. The matrix representation of the $CR_Y(\theta)$ can be written as 

\begin{equation}
    CR_Y(\theta)=
    \begin{bmatrix}
    {1} & \hspace{4mm}{0} & {0} & {0} \\
    {0} & \hspace{4mm}{1} & {0} & {0} \\
    {0} & \hspace{4mm}{0} & \cos\Big(\dfrac{\theta}{2}\Big)& - \sin\Big(\dfrac{\theta}{2}\Big) \\[2mm]
    {0} & \hspace{4mm} {0} & \sin\Big(\dfrac{\theta}{2}\Big)& \cos\Big(\dfrac{\theta}{2}\Big) \\
    \end{bmatrix}
\end{equation}

\subsubsection{Three-qubit gates} 
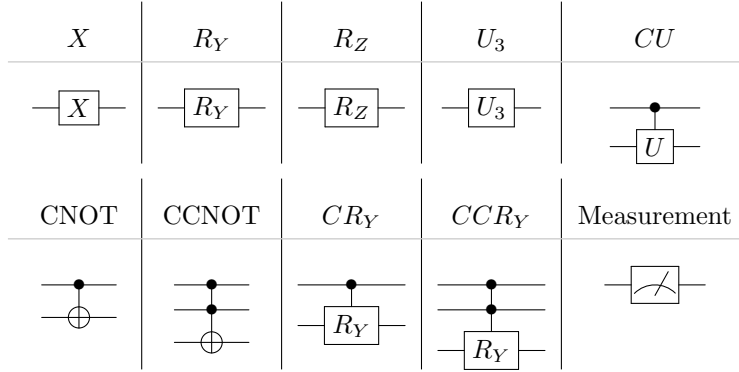
\begin{figure}[!h]
\centering
\begin{tabular}{c|c|c|c|c}

\rule{0pt}{4ex} 
$X$  & $R_Y$ & $R_Z$ & $U_3$ & $CU$ \\
\arrayrulecolor{lightgray}
\hline
\arrayrulecolor{black}
\rule{0pt}{4ex} 
\Qcircuit @C=1em @R=.7em{& \gate{X} & \qw}    &  \Qcircuit @C=1em @R=.7em { & \gate{R_Y} & \qw} & \Qcircuit @C=1em @R=.7em {& \gate{R_Z} & \qw} & \Qcircuit @C=1em @R=.7em {& \gate{U_3} & \qw} &\Qcircuit @C=1em @R=.7em {& \ctrl{1} & \qw\\& \gate{U} & \qw} \vspace{0.2cm}\\ 

\rule{0pt}{4ex} 
CNOT  & CCNOT & $CR_Y$ & $CCR_Y$  & Measurement \\
\arrayrulecolor{lightgray}
\hline
\arrayrulecolor{black}
\rule{0pt}{4ex} 
\Qcircuit @C=1em @R=.7em {& \ctrl{1} & \qw\\& \targ & \qw}   & 
\Qcircuit @C=1em @R=.7em {& \ctrl{1} & \qw \\& \ctrl{1} & \qw \\& \targ   & \qw} &
\Qcircuit @C=1em @R=.7em {& \ctrl{1} & \qw \\& \gate{R_Y} & \qw } &
\Qcircuit @C=1em @R=.7em {& \ctrl{1} & \qw \\& \ctrl{1} & \qw \\& \gate{R_Y}  & \qw} \vspace{0.2 cm} &
\Qcircuit @C=1em @R=.7em {& \meter & \qw } \\
\end{tabular}
\caption{Representation of commonly used one, two, and three qubit gates}  
\label{fig:gates}
\end{figure}

We will discuss two three-qubit gates that are later used in the proposed methodology: CCNOT (or $CCX$ or Toffoli) and $CCR_Y(\theta)$. Out of the three qubits on which the CCNOT and $CCR_Y(\theta)$ are implemented, two qubits act as control qubits and the other is the target qubit. In the case of CCNOT gate, when both control qubits are $\Ket{1}$, we implement the $X$ gate on the target qubit. In the case of the $CCR_Y(\theta)$ gate, we implement the $R_Y(\theta)$ gate when the two control qubits are in the $\Ket{1}$ state. The three-qubits are not elementary gates but can be decomposed into a combination of single-qubit and CNOT gates. For example, the CCNOT can be represented using a combination of nine single qubit gates and six CNOT gates \citep{shende2008cnot}. Similar to $CCX$ and $CCR_Y(\theta)$, we can define an $C^nX$ and $C^nR_Y(\theta)$ gates with $n$ control qubits and one target qubits \mbox{\citep{liu2008analytic}}. Figure \ref{fig:gates} provides the representation of various gates discussed above in a quantum circuit. A quantum circuit represents a graphical representation of a sequence of gates that are implemented on various qubits to obtain a desired computation. Measurement gate in Figure \ref{fig:gates} performs the measurement operation on a qubit.

\begin{figure}
\begin{center}

\hspace{0.1mm} \Qcircuit @C=1em @R=0.9em {
     & \ctrl{1} & \qw&  \\
     & \gate{U} & \qw & \push{=}} \hspace{5mm} \Qcircuit @C=1em @R=0.9em {
    & \qw & \ctrl{1} & \qw & \ctrl{1} & \qw& \qw &\\
    & \gate{A} & \targ & \gate{B} & \targ & \gate{C}& \qw
 }

\vspace{4mm}
\caption{Decomposition of a CU gate into a combination of single qubit and CNOT gates}
\label{fig:decomposeU}
\end{center}

\end{figure}
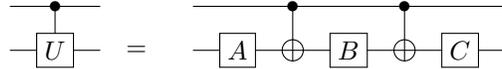


According to \cite{nielsen2002quantum}, the $CU$ can be decomposed into a combination of single-qubit and CNOT gates as given in Figure \ref{fig:decomposeU}; this decomposition can mathematically be represented as

\begin{equation}
\label{eqn:cu}
CU = (I\otimes A)CX(I\otimes B)CX(I\otimes C)
\end{equation}

\noindent where $A = R_Z(\phi)R_Y(\theta/2)$, $B = R_Y(-\theta/2)R_Z(-(\lambda+\phi)/2)$, $C = R_Z((\lambda-\phi)/2)$, and $I$ represents the identity matrix. $I\otimes A$ represents the tensor product of two matrices, $I$ and $A$, where $I$ and $A$ are single-qubit gates implemented on the first and second qubits respectively. $I\otimes A$ is the simplified representation of the two gates acting on the two qubits.


\subsection{Bayesian Networks}
\label{subsec:BN}

As mentioned in Section \ref{sec:introduction}, Bayesian networks (BNs) are probabilistic graphical models that represent a probabilistic framework to model stochastic/uncertain systems. In a probabilistic framework, a stochastic system can be represented as a joint probability distribution defined over the set of random variables. A BN consists of nodes and edges, where nodes represent the random variables, and edges represent the dependence between nodes, which is quantified using conditional probability tables (CPT, for discrete variables) and conditional probability distributions (CPD, for continuous variables). In this paper, we consider the design of quantum circuits to represent discrete Bayesian networks \mbox{\citep{dash2004model}}. 

Let us consider a Bayesian network with $s$ nodes, where $\mathbb{V}=\{V_1,V_2,...,V_s\}$ represents the set of all nodes or random variables. An edge from node $V_i$ to node $V_j$ represents the dependence between the variables $V_i$ and $V_j$, and that the values of $V_j$ are dependent on the values of $V_i$. Here, $V_i$ is called the parent node and node $V_j$ is referred to as the child node. The nodes without any parent nodes are typically referred to as root nodes. Using the graphical representation of a Bayesian network, the joint probability over the nodes (random variables) can be decomposed into a product of marginal and conditional probabilities as \mbox{\citep{koller2009probabilistic}}
 
\begin{equation}
P(V_1,V_2,...,V_s)= \prod_{i=1}^{s} P(V_i| \Pi_{V_i})
\end{equation}
where $\Pi_{V_i}$ denotes the set of parents nodes associated with $V_i$. For root nodes, the $P(V_i| \Pi_{V_i})$ becomes equal to the marginal distribution, $P(V_i)$. 

\begin{figure}
    \centering
    \includegraphics[width=0.98\columnwidth]{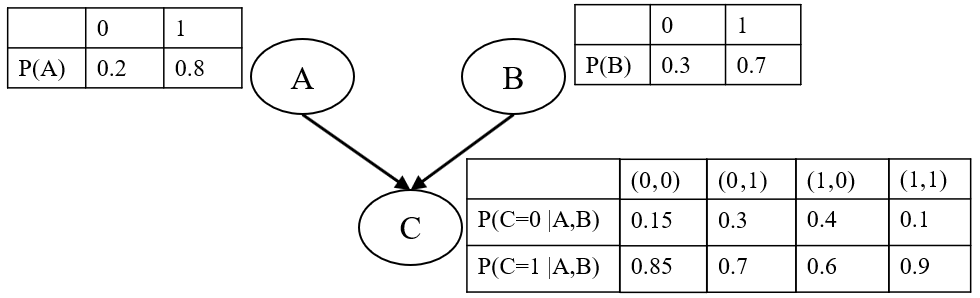}
    \caption{An illustrative 3-node Bayesian network with binary variables}
    \label{fig:simpleBN}
\end{figure}

Consider a simple Bayesian network with 3 nodes as shown in Figure \ref{fig:simpleBN} where A, B and C are discrete random variables with two states \textit{True} (or ``1") and \textit{False} (or ``0"). In this BN, $A$ and $B$ are root nodes whereas $C$ is a child node with $A$ and $B$ as parent nodes. We have the marginal distributions for root nodes (shown in Figure \ref{fig:simpleBN} and a CPT for the child node, which represents a probability distribution of the child node conditioned on the values taken by the parent nodes. Given a CPT, we can calculate its marginal distribution by integrating over the distributions of the parent nodes, i.e., $P(C) = \sum_{A, B} P(C|A, B)\times P(A, B)$. Due to the independence between nodes A and B (from Figure \ref{fig:simpleBN}), the joint probabilities of A and B can be written as the product of individual probabilities, i.e., $P(A, B) = P(A)\times P(B)$, and therefore, $P(C) = P(C|A, B)\times P(A)\times P(B)$ \mbox{\citep{koller2009probabilistic}}.

We use the marginal probabilities of various nodes in a Bayesian network as a measure to check the accuracy of the proposed quantum circuit representation approach in Section \ref{sec:method}. After we design the quantum circuit, we estimate the marginal probabilities by simulating the quantum circuits, and compare them against the values from classical Bayesian network implementations. After providing a brief background to quantum computing and Bayesian networks, we will now discuss the representation of a Bayesian network through a quantum circuit.

\section{Quantum circuit of a Bayesian network}
\label{sec:method}

We use the following principles for the design of a quantum circuit to represent a Bayesian network.

\begin{enumerate}[label=(\roman*)]
    \item Map each node in a Bayesian network to one or more qubits (depending on the number of discrete states of the node)
    \item Map the marginal/conditional probabilities of each node to the probability amplitudes (or probabilities) associated with various states of the qubit(s).
    \item Realize the required probability amplitudes of quantum states using (controlled) rotation gates.
\end{enumerate}

Let us discuss how these ideas can be used to design a quantum circuit for the 3-node Bayesian network in Figure \ref{fig:simpleBN}. All the nodes in Figure \ref{fig:simpleBN} have two states (0 and 1), and since a qubit can represent two states ($\Ket{0}$ and $\Ket{1}$), we can map each node to a different qubit. Also, let us map state 0 of each node to $\Ket{0}$ and state 1 to $\Ket{1}$. Let the three nodes A, B, and C be mapped to three qubits $q_0, q_1,$ and $q_2$ respectively.  Let the initial state of the three qubits be $\Ket{0}$. In the cases of $q_0$ and $q_1$, we will need to apply rotation gates ($R_Y$) with angles ($\theta_A$ and $\theta_B$) that result in superposed quantum states whose probabilities correspond the probabilities of nodes A and B respectively. The calculation of those angles are later discussed in Section \ref{subsec:root_twonode}. In the case of qubit $q_2$ (that corresponds to C), we will have a different rotation angle conditioned on the states of $q_0$ and $q_1$ since we have a different set of probabilities for node C conditioned on the values of parent nodes (A, B). Since there are four combinations of parent node values, we will have four rotation angles, one for each parent node combination. These conditional rotations are implemented using controlled rotation gates, whose angles depend on the conditional probabilities of C; these rotations are represented as $\theta_{C,ij}$, where $i,j=0,1$ and represent the states of $q_0$ and $q_1$ respectively.


\begin{figure}[!h]
\begin{center}
\scalebox{0.75}{
\hspace*{12mm} \Qcircuit @C=0.5em @R=0.9em {
 && & & \mbox{$\ket{00}$} & && & \mbox{$\ket{01}$}& && & \mbox{$\ket{10}$} & && & \mbox{$\ket{11}$} & & & &  \\
\lstick{q_0:\Ket{0}}  &\gate{R_Y(\theta_A)}\barrier[0em]{0}&\qw &\gate{X}&  \ctrl{1} &\gate{X} \barrier[0em]{0}&\qw&\gate{X}& \ctrl{1} &\gate{X} \barrier[0em]{0}&\qw&\qw& \ctrl{1} &\qw \barrier[0em]{0}&\qw&\qw&  \ctrl{1} &\qw& \qw  \\
\lstick{q_1:\Ket{0}} &\gate{R_Y(\theta_B)}\barrier[0em]{0}&\qw &\gate{X}&  \ctrl{1} &\gate{X} \barrier[0em]{0}&\qw&\qw&  \ctrl{1} &\qw \barrier[0em]{0}&\qw&\gate{X}&  \ctrl{1} &\gate{X} \barrier[0em]{0}&\qw&\qw&  \ctrl{1} &\qw& \qw  \\
\lstick{q_2:\Ket{0}} &\qw \barrier[0em]{0}&\qw &\qw& \gate{R_Y(\theta_{C,00})} &\qw \barrier[0em]{0}&\qw&\qw&  \gate{R_Y(\theta_{C,01})} &\qw\barrier[0em]{0}&\qw&\qw& \gate{R_Y(\theta_{C,10})} &\qw \barrier[0em]{0}&\qw&\qw&  \gate{R_Y(\theta_{C,11})} &\qw&\qw }}
\end{center}

\caption{Illustrative quantum circuit with three qubits for the 3-node Bayesian network in Figure 4}
\label{fig:example}
\end{figure}
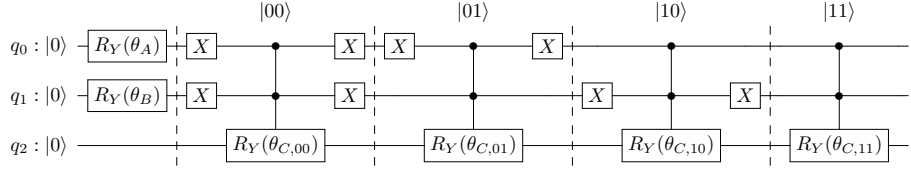

Figure \ref{fig:example} provides a conceptual quantum circuit for the 3-node Bayesian network. First, we implement the single qubit rotations to obtain the probabilities associated with A and B. Depending on the values of A and B, we implement controlled rotations to realize the conditional probabilities associated with C. In a controlled rotation, since the rotations are applied only when the control qubit is $\Ket{1}$, we use the $X$ gate to flip the $\Ket{0}$ to $\Ket{1}$ state to obtain the conditional probabilities when the parent node value(s) are 0. As there are two parent nodes, we implement a $CCR_Y$ gate to realize the conditional probabilities of C for every combination of the parent nodes. If there are $n$ parent nodes, then we would implement a $C^nR_Y$ gate. The overall quantum circuit can be obtained by composing all the single-qubit and controlled rotations in a sequential manner (as shown in Figure \ref{fig:example}). In this paper, we refer to this approach as the C-QBN approach, which stands for Compositional approach for Quantum Bayesian networks.

We discuss below the computation of rotation angles to realize nodal probabilities in Section \ref{subsec:root_twonode}, representation of two-state child nodes with one or more parent nodes in Section \ref{subsec:2node_child}, and representation of nodes with more than two states in Section \ref{subsec:multistate}.

\subsection{Rotation angle computation}
\label{subsec:root_twonode}

As mentioned above, we can represent a two-state root node using a single qubit. By applying an $R_Y$ gate with an appropriate angle, the probabilities of the root node can be mapped to the probabilities (and thus probability amplitudes) of the basis states, $\Ket{0}$ and $\Ket{1}$. Let $\theta_{V_i}$ represent the rotation angle associated with a two-state root node, $V_i$. Given the initial state of a qubit as $\Ket{0}$, the application of $R_Y(\theta)$ will transform $\Ket{0}$ to $\cos{\bigg(\dfrac{\theta}{2}\bigg)}\Ket{0} + \sin{\bigg(\dfrac{\theta}{2}\bigg)}\Ket{1}$. Therefore, the probabilities associated with the $\Ket{0}$ and $\Ket{1}$ states are equal to $\cos^2{\bigg(\dfrac{\theta}{2}\bigg)}$ and $\sin^2{\bigg(\dfrac{\theta}{2}\bigg)}$ respectively. If $P(V_i=0)$ and $P(V_i=1)$ represent the probabilities of states 0 and 1 of $V_i$, then the rotation angle can be computed as 

\begin{equation}
\label{eqn:theta}
    \theta_{V_i} = 2\times\tan^{-1}\sqrt{\dfrac{P(\Ket{1})}{P(\Ket{0})}} = 2\times\tan^{-1}\sqrt{\dfrac{P(V_i=1)}{P(V_i=0)}}
\end{equation}

In Eq. (\ref{eqn:theta}), $P(\Ket{0})$ and $P(\Ket{1})$ represent the probabilities of a qubit to be in $\Ket{0}$ and $\Ket{1}$ respectively. Since we map the nodal probabilities to the probabilities of quantum states, $P(\Ket{0})$ and $P(\Ket{1})$ are replaced with $P(V_i=0)$ and $P(V_i=1)$ respectively. Therefore, two-state root nodes can be represented using an $R_Y$ gate with a rotation angle of $2\times\tan^{-1}\sqrt{\dfrac{P(V_i=1)}{P(V_i=0)}}$. In Figure \ref{fig:simpleBN}, the rotation angle to find the probabilities of A and B can be calculated as $\theta_A = 2\times\tan^{-1}\sqrt{\dfrac{0.8}{0.2}} = 2.214$ and $\theta_B = 2\times\tan^{-1}\sqrt{\dfrac{0.7}{0.3}} = 1.982$. 

Eq. (\ref{eqn:theta}) can also be used to compute the rotation angles associated with conditional probabilities of two-state child nodes. Let $V_i$ and $\Pi_{V_i}$ represent a child node and set of its parent nodes respectively. For each combination of parent node values, $\Pi_{V_i} = \Pi^*_{V_i}$, we have probabilities for $V_i=0$ and $V_i=1$ denoted as $P(V_i=0|\Pi_{V_i} = \Pi^*_{V_i})$ and $P(V_i=1|\Pi_{V_i} = \Pi^*_{V_i})$ respectively. The rotation angle associated with $V_i$ when $\Pi_{V_i} = \Pi^*_{V_i}$, which is denoted by $\theta_{V_i, \Pi^*_{V_i}}$ can be calculated as 

\begin{equation}
\label{eqn:theta2}
    \theta_{V_i, \Pi^*_{V_i}} = 2\times\tan^{-1}\sqrt{\dfrac{P(V_i=1|\Pi_{V_i} = \Pi^*_{V_i})}{P(V_i=0|\Pi_{V_i} = \Pi^*_{V_i})}}
\end{equation}

The rotation angles associated with node C (qubit $q_2$ in Figure \ref{fig:example}) can be calculated using Eq. (\ref{eqn:theta2}) as $\theta_{C,00} = 2\times\tan^{-1}\sqrt{\dfrac{0.85}{0.15}} = 2.346$, $\theta_{C,01} = 2\times\tan^{-1}\sqrt{\dfrac{0.7}{0.3}} = 1.982 $, $\theta_{C,10} = 2\times\tan^{-1}\sqrt{\dfrac{0.6}{0.4}} = 1.772 $, and $\theta_{C,11} = 2\times\tan^{-1}\sqrt{\dfrac{0.9}{0.1}} = 2.498$. The conditional probabilities associated with child nodes are realized through controlled rotations. As controlled rotation gates are not elementary gates, they need to be decomposed into single-qubit and two-qubit elementary gates; the decomposition is discussed below in Section \ref{subsec:2node_child}.

\subsection{Representing two-state child nodes with two-state parent nodes}
\label{subsec:2node_child}

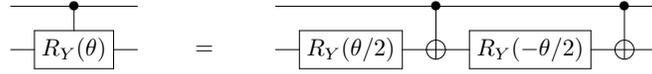
\begin{figure}[H]
\begin{center}
\scalebox{0.9}{

\hspace{1mm} \Qcircuit @C=1em @R=0.9em {
     & \ctrl{1} & \qw\\
     & \gate{R_Y (\theta)} & \qw & \push{\hspace{5mm}=}} \hspace{8mm} \Qcircuit @C=1em @R=0.9em {
    & \qw & \ctrl{1} & \qw & \ctrl{1} & \qw\\
    & \gate{R_Y(\theta/2)} & \targ & \gate{R_Y(-\theta/2)} & \targ & \qw}}
\end{center}
\vspace{4mm}
\caption{Decomposition of a $CR_Y$ gate into single-qubit $R_y$ and CNOT gates}
\label{fig:decomposeCRY}
\end{figure}

First, we consider the representation of child nodes with one parent node, and then we consider the case with multiple parent nodes. A two-state child node with one parent node can be represented using two $CR_Y$ gates (assuming the parent node has two states) with rotation angles computed using Eq. (\ref{eqn:theta2}) conditioned on the parent node value. We consider parent nodes with more than two states in Section \ref{subsec:multistate}. The $CR_Y$ gate is a special case of a $CU$ gate discussed in Section \ref{subsec:gates}, where $U=R_Y$. In Section \ref{subsec:gates}, we discussed the decomposition of a $CU$ gate in terms of elementary single-qubit and CNOT gates shown in Figure \ref{fig:decomposeU} and given in Eq. (\ref{eqn:cu}). Since $R_Y(\theta) = U_3\Big(\dfrac{\theta}{2},0,0\Big)$, the single qubit gates A, B, and C in  Figure \ref{fig:decomposeU} can be calculated as $A = R_Z(0)R_Y\Big(\dfrac{\theta}{2}\Big) = R_Y\Big(\dfrac{\theta}{2}\Big), B = R_Y\Big(-\dfrac{\theta}{2}\Big)R_Z\Big(-\dfrac{(0+0)}{2}\Big) =  R_Y\Big(-\dfrac{\theta}{2}\Big),$ and $C = R_Z\Big(\dfrac{(0-0)}{2}\Big) = I$. Figure \ref{fig:decomposeCRY} illustrates the decomposition of the $CR_Y$ gate into single-qubit and CNOT gates. After considering child nodes with one parent node, we now consider child nodes with more than one parent nodes.

\begin{figure}[!h]
\begin{center}
\scalebox{0.85}{
\hspace{7mm} \Qcircuit @C=0.5em @R=0.9em {
    \lstick{q_0:\Ket{0}} & \ctrl{1} &\qw   \\
    \lstick{q_1:\Ket{0}}  & \ctrl{1}&\qw  \\
    \lstick{q_2:\Ket{0}}  & \ctrl{1}&\qw  \\
    \lstick{q_3:\Ket{0}}  & \ctrl{1}&\qw &\push{=}   \\
    \lstick{q_4:\Ket{0}}  & \ctrl{1} &\qw \\
    \lstick{q_{5}:\Ket{0}}  & \gate{R_Y (\theta)}&\qw } \hspace{15mm} \Qcircuit @C=0.6em @R=0.9em {
    \lstick{q_0:\Ket{0}}    & \ctrl{1}  &\qw   & \qw & \qw &\qw& \qw&\qw&\qw&\qw&\qw&\qw&\ctrl{1}&\qw\\
    \lstick{q_1:\Ket{0}}    & \ctrl{4}  &\qw &\qw &\qw & \qw& \qw& \qw&\qw&\qw&\qw&\qw&\ctrl{4}&\qw\\
    \lstick{q_2:\Ket{0}}    & \qw       &\ctrl{3} & \qw &\qw& \qw& \qw& \qw&\qw&\qw&\qw&\ctrl{3}&\qw &\qw\\
    \lstick{q_3:\Ket{0}}    & \qw   &\qw & \ctrl{3}  &\qw& \qw& \qw& \qw&\qw&\qw&\ctrl{3}&\qw&\qw&\qw\\
    \lstick{q_4:\Ket{0}}    & \qw   &\qw  & \qw & \ctrl{3}& \qw& \qw& \qw&\qw&\ctrl{3}&\qw&\qw&\qw&\qw\\
    \lstick{a_{0}:\Ket{0}}  & \targ &\ctrl{1} & \qw &\qw& \qw& \qw& \qw&\qw&\qw&\qw&\ctrl{1}&\targ&\qw\\
    \lstick{a_{1}:\Ket{0}}  & \qw   &\targ & \ctrl{1}&\qw& \qw& \qw& \qw&\qw&\qw&\ctrl{1}&\targ&\qw&\qw\\
    \lstick{a_{2}:\Ket{0}}  & \qw   &\qw & \targ &\ctrl{1}& \qw& \qw& \qw&\qw&\ctrl{1}&\targ&\qw&\qw&\qw\\
    \lstick{a_{3}:\Ket{0}}  & \qw   &\qw & \qw &\targ& \qw& \ctrl{1}& \qw&\ctrl{1}&\targ&\qw&\qw&\qw&\qw\\
    \lstick{q_{5}:\Ket{0}}  & \qw   &\qw & \qw &\qw& \gate{R_Y(\theta/2)}&\targ&\gate{R_Y(-\theta/2)}&\targ&\qw&\qw&\qw&\qw&\qw\\
    }}
\end{center}
\vspace{4mm}
\caption{Representation of $C^5R_Y$ gate using four ancilla qubits, single qubit $R_Y$, CNOT and CCNOT gates}
\label{fig:decompose_CnRY}
\end{figure}
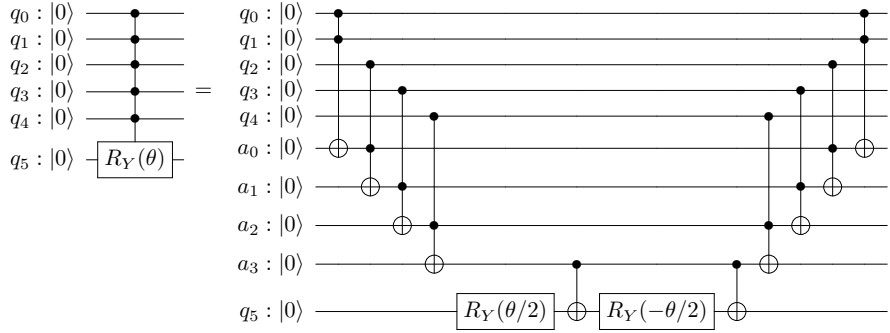

Let $n$ represent the number of parent nodes for a child node, $V_i$. The conditional probabilities of $V_i$ can be stated using $C^nR_Y$ gate where the $n$ control qubits are the $n$ qubits corresponding to the $n$ parent nodes and the target qubit represents the child node.  For the child node C in Figure \ref{fig:simpleBN}, n=2, and therefore, we used a $CCR_Y$ or $C^2R_Y$ gate to show the conditional probabilities of C in Figure \ref{fig:example}. $C^nR_Y$ is not an elementary gate and will need to be decomposed into elementary gates. One of the techniques to build the $C^nR_Y$ is the use of additional ``dummy" qubits known as ancilla qubits \citep{nielsen2002quantum}. Following \citep{nielsen2002quantum}, implementation $C^nR_Y$ requires $n-1$ ancilla qubits. Using ancilla qubits, the $C^nR_Y$ gate is decomposed into a combination of $2(n-1)$ CCNOT gates, and one $CR_Y$ gate. For illustration, Figure \ref{fig:decompose_CnRY} details the representation of $C^5R_Y$ gate using four ancilla qubits. The $CR_Y$ gate can again be decomposed into a combination of single qubit and CNOT gates as detailed in Figure \ref{fig:decomposeCRY}.

In Figure \ref{fig:decompose_CnRY}, $q_i, i=0\dots 4$ represent the control qubits (parent nodes), and $q_5$ is the target qubit (child node). As there are five control qubits, we use four ancilla qubits ($a_j, j=0\dots 3$). In total, we have 8 CCNOT gates ($2\times (5-1)$), two CNOTs and two single-qubit rotation gates. In Figure \ref{fig:example}, there are two control qubits ($n=2$) for $q_2$ (node C), we will use one ancilla qubit to make the $CCR_Y$ gates. Note that we do not need a different set of ancilla qubits for implementing various controlled rotations (conditional probabilities), and we can the same set of ancilla qubits for all the $C^nR_Y$ rotations. Moreover, we can use the same set of ancilla qubits to build the $C^nR_Y$ rotations associated with various child nodes. Consider a Bayesian network with $s$ two-state nodes given by ${V_1, V_2,\dots V_s}$ and let $|.|$ denote the cardinality operator. For a node $V_i$, $|\Pi_{V_i}|$ provides the number of parent nodes of a $V_i$. For a root node,  $|\Pi_{V_i}| = 0$ and for a child node, $|\Pi_{V_i}| > 0$. Therefore, the total number of qubits required to represent a Bayesian network with two-state nodes (denoted as $m_{BN,2}$) can be calculated as 

\begin{equation}
\label{eqn:total_2qubit}
m_{BN,2} = s + \max \Big(|\Pi_{V_1}|, |\Pi_{V_2}|, \dots |\Pi_{V_s}|\Big) - 1
\end{equation}

\noindent where $s$ qubits are used to represent $s$ nodes in the Bayesian network, and an additional $\max \Big(|\Pi_{V_1}|, |\Pi_{V_2}|, \dots |\Pi_{V_s}|\Big)-1$ ancilla qubits are used to represent the multi-qubit conditional rotations. 


\subsection{Representing discrete variables with more than two states}
\label{subsec:multistate}
 When a node has more than two states, then we need to use more than one qubit to represent it, as one qubit can represent only two states. Consider a random variable $V_i$ with $n_i$ states, denoted as $V_{i,j}, j=0,1,\dots n_i-1$, and $P(V_{i,j})$ represents the probability of state $V_{i,j}$; therefore, $\sum_{j=0}^{n_i-1} P(V_{i,j})=1$. 
 
 Let $m_i$ represent the number of qubits to represent $V_i$. The number of states that are represented by $m_i$ qubits is $2^{m_i}$, which needs to be greater than or equal to $n_i$. Thus, value of $m_i$ can be calculated as the smallest integer that is greater than or equal to $\log_2 n_i$, which can be represented using the ceiling function as $m_i = \ceil[\big]{\log_2 n_i}$, where $\ceil[\big]{.}$ is the ceiling function. Let $\Ket{q_j} j=l\dots l+m_i-1$ represent $m_i$ qubits in the quantum circuit used to represent $V_i$. In addition to qubits representing node $V_i$, there could be other qubits in the quantum circuit, which are used to represent other nodes and/or ancilla qubits. Here, $l$ is used to represent the indices of the qubits used to represent node $V_i$. The superposition state, $\Ket{q_l q_{l+1} \dots q_{l+m_i-1}}$ can be written in terms of the basis states as 
 
\begin{equation}
\label{eqn:mstate}
\Ket{q_l q_{l+1} \dots q_{l+m_i-1}} = \sum_{j=0}^{2^{m_i}-1} \alpha_t \Ket{q_l q_{l+1} \dots q_{l+m_i-1}}_j
\end{equation}

\begin{figure}[!h]
\begin{center}
\hspace{1mm}\Qcircuit @C=1em @R=0em {
\lstick{\Ket{0}}& \qw & \multigate{4}{U_{0,1 \dots m_i-1}} & \qw & \qw \\
\lstick{\Ket{0}}& \qw & \ghost{U_{0,1 \dots m_i-1}} & \qw & \qw \\
\lstick{\Ket{0}}& \qw & \ghost{U_{0,1 \dots m_i-1}} & \qw & \qw  & \push{=} \\
& \cdots & \nghost{U_{0,1 \dots m_i-1}} & \cdots & \\
\lstick{\Ket{0}}& \qw    & \ghost{U_{0,1 \dots m_i-1}} & \qw & \qw}  \hspace{10mm}\Qcircuit @C=1em @R=0em {
\lstick{\Ket{0}}& \gate{R_Y(\theta_l)} & \ctrl{1} & \gate{X} & \ctrl{1} & \gate{X} & \qw \\
\lstick{\Ket{0}}& \qw & \multigate{3}{U_{1 \dots m_i-1, q_l = \Ket{1}}} & \qw & \multigate{3}{U_{1 \dots m_i-1, q_l = \Ket{0}}} &\qw & \qw\\
\lstick{\Ket{0}}& \qw & \ghost{U_{1 \dots m_i-1, q_l = \Ket{1}}} & \qw & \ghost{U_{1 \dots m_i-1, q_l = \Ket{0}}}&\qw& \qw \\
& \cdots & \nghost{U_{1 \dots m_i-1, q_l = \Ket{1}}} & \cdots & \nghost{U_{1 \dots m_i-1, q_l = \Ket{0}}} & \cdots\\
\lstick{\Ket{0}}& \qw & \ghost{U_{1 \dots m_i-1, q_l = \Ket{1}}} & \qw & \ghost{U_{1 \dots m_i-1, q_l = \Ket{0}}}&\qw& \qw}
\end{center}
\vspace{4mm}
\caption{Decomposing a $m_i$ qubit rotation into single qubit  and Controlled $m_i$ - 1 qubit rotations}
\label{fig:kqdecompose}
\end{figure}
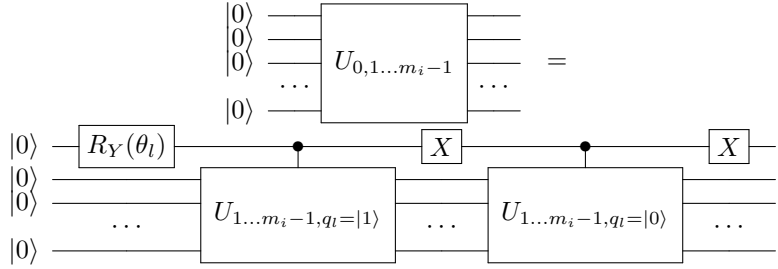

In Eq. (\ref{eqn:mstate}), $\Ket{q_l q_{l+1} \dots q_{l+k-1}}_j$ represents a basis state ($\Ket{00\dots 0}$ for instance) and $\alpha_j$ represents its probability amplitude. Let us map the $n_i$ states of the variable to $n_i$ basis states represented by these $m_i$ qubits. Hence, state $V_{i,j}$ can be mapped to the quantum state $\Ket{q_l q_{l+1} \dots q_{l+m_i-1}}_j$. The probability amplitudes of these $n_i$ quantum states can be calculated using the available state probabilities and the probability amplitudes of the remaining quantum states are set to zero. Therefore, $\alpha_j = \sqrt{P(V_{i,j})}$ when $j<n_i$ and $\alpha_j = 0$ when $n_i\leq j < 2^{m_i}$. 

Our goal is to identify the gate $U$ that acts on $m_i$ qubits and produces the desired state probabilities, i.e., $U\Ket{0}^{\otimes m_i} = \Ket{q_l q_{l+1} \dots q_{l+m_i-1}}$ which is equlal to $\sum_{j=0}^{n_i-1} \sqrt{P(V_{i,j})} \Ket{q_l q_{l+1} \dots q_{l+m_i-1}}_j$. Since $U$ is a multi-qubit gate, it needs to be decomposed into a set of elementary gates discussed in Section \ref{sec:background}. Our approach is to decompose the probability distribution defined over $m_i$ qubits into a combinations of marginal and conditional distributions that can be implemented using single-qubit and controlled-rotations. First, we rotate the first qubit ($q_l$) to obtain its associated probability values corresponding to its $\Ket{0}$ and $\Ket{1}$ states respectively using a single-qubit $R_Y$ rotation, and we implement a different multi-qubit rotation on the remaining $k-1$ when $q_l=\Ket{0}$ and $q_l=\Ket{1}$.

Figure \ref{fig:kqdecompose} details the decomposition of a $m_i$ qubit rotation into a combination of single-qubit and controlled $m_i$-1 qubit rotations, where $U_{0,1 \dots m_i-1}$ is the $m_i$ qubit rotation, and $U_{1 \dots m_i-1, q_l=\Ket{1}}$ and $U_{1 \dots m_i-1, q_1=\Ket{0}}$ are the rotations implemented on $m_i-1$ qubits ($q_{l+1},q_{l+2} \dots q_{l+m_i-1}$) when $q_l=\Ket{1}$ and $q_l=\Ket{0}$ respectively. $R_Y(\theta_l)$ represents the rotation to obtain the probabilities associated with qubit $q_l$, and can be calculated using Eq. (\ref{eqn:theta}). The probabilities of $\Ket{0}$ and $\Ket{1}$ states of $q_l$ can be calculated using an indicator function, $\mathbb{I}_{q_l}$, defined as 

\begin{equation}
\label{eqn:ql}
\mathbb{I}_{q_l}(\Ket{q_l q_{l+1} \dots q_{l+m_i-1}})=
\left\{
	\begin{array}{ll}
		1  & \mbox{if } \Ket{q_l}=\Ket{1} \\
		0 & \mbox{if } \Ket{q_l}=\Ket{0}
	\end{array}
\right.
\end{equation}

Using Eq. (\ref{eqn:ql}), the probability that $q_l=\Ket{1}$ can be calculated as $P(q_l=\Ket{1}) = \sum_{j=0}^{2^{m_i}-1} \alpha_j^2 \mathbb{I}_{q_l}(\Ket{q_l q_{l+1} \dots q_{l+m_i-1}}_j)$, and $P(q_l=\Ket{0}) = 1-P(q_l=\Ket{1})$. The rotation angle, $\theta_l$ in Figure \ref{fig:kqdecompose} can be calculated using Eq. (\ref{eqn:theta}) as 

\begin{equation}
\label{eqn:rot_ql}
\begin{aligned}
\theta_l & = 2\times \tan^{-1} \sqrt{\dfrac{P(q_l=\Ket{1})}{P(q_l=\Ket{0})}} \\
 & = 2\times \tan^{-1}\sqrt{\dfrac{\sum_{j=0}^{2^{m_i}-1} \alpha_j^2 \mathbb{I}_{q_l}(\Ket{q_l q_{l+1} \dots q_{l+m_i-1}}_i)}{1-\sum_{j=0}^{2^{m_i}-1} \alpha_j^2 \mathbb{I}_{q_l}(\Ket{q_l q_{l+1} \dots q_{l+m_i-1}}_j)}}
\end{aligned}
\end{equation}

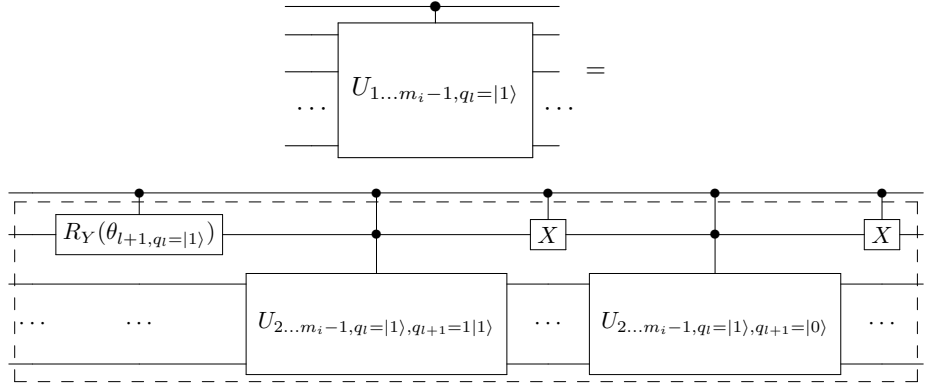
\begin{figure}[!h]
\begin{center}

\hspace{1mm} \Qcircuit @C=1em @R=0.5em  {
&\qw& \ctrl{1} & \qw \\
&\qw& \multigate{3}{U_{1 \dots m_i-1, q_l=\Ket{1}}}& \qw\\
&\qw& \ghost{U_{1 \dots m_i-1, q_l=\Ket{1}}}& \qw & \push{=} \\
&\cdots& \nghost{U_{1 \dots m_i-1, q_l=\Ket{1}}} & \cdots \\
&\qw& \ghost{U_{1 \dots m_i-1, q_l=\Ket{1}}}&\qw} \hspace{4mm}
\vspace{4mm}

\scalebox{0.89}{
\Qcircuit @C=1em @R=0.8em {
&\qw & \ctrl{1} & \ctrl{1}&\ctrl{1}&\ctrl{1} & \ctrl{1} & \qw\\
&\qw & \gate{R_Y(\theta_{l+1,q_l=\Ket{1}})} & \ctrl{1}&\gate{X}&\ctrl{1} & \gate{X} &\qw \\ 
&\qw& \qw & \multigate{2}{U_{2 \dots m_i-1, q_l=\Ket{1}, q_{l+1}=1\Ket{1}}}&\qw & \multigate{2}{U_{2 \dots m_i-1, q_l=\Ket{1}, q_{l+1}=\Ket{0}}}&\qw&\qw\\
&\cdots & \cdots & \nghost{U_{2 \dots m_i-1, q_l=\Ket{1}, q_{l+1}=1\Ket{1}}}&\cdots  & \nghost{U_{2 \dots m_i-1, q_l=\Ket{1}, q_{l+1}=\Ket{0}}} & \cdots\\
&\qw &\qw&\ghost{U_{2 \dots m_i-1, q_l=\Ket{1}, q_{l+1}=1\Ket{1}}}&\qw & \ghost{U_{2 \dots m_i-1, q_l=\Ket{1}, q_{l+1}=\Ket{0}}} & \qw&\qw
\gategroup{2}{2}{5}{7}{1.5em}{--}}}
\end{center}
\vspace{4mm}
\caption{Decomposing a controlled $m_i$-1 qubit rotation into controlled single qubit  and controlled-controlled $m_i$-2 qubit rotations}
\label{fig:k1qdecompose}
\end{figure}

The above procedure for decomposing $m_i$ qubit rotation into a single qubit and controlled $m_i$-1 qubit rotations can again be used to decompose the controlled $m_i$-1 qubit rotation resulting in controlled single-qubit and controlled-controlled $m_i$-2 qubit rotations. We will illustrate the decomposition of $CU_{1 \dots m_i-1, q_l=\Ket{1}}$ in Figure \ref{fig:k1qdecompose}. 

First, we implement a $CR_Y$ gate on qubit $q_{l+1}$ to obtain the probabilities of $q_{l+1}$ when $q_l=\Ket{1}$. The probabilities of $q_{l+1}=\Ket{0}$ and $q_{l+1}=\Ket{1}$ when $q_l=\Ket{1}$ are calculated using another indicator function defined over $q_l$ and $q_{l+1}$ as

\begin{equation}
\label{eqn:ql1}
\begin{split}
&\mathbb{I}_{q_l=1, q_{l+1}}(\Ket{q_l q_{l+1} \dots q_{l+m_i-1}})\\
&= \left\{
	\begin{array}{ll}
		1  & \mbox{if } \Ket{q_l}=\Ket{1} \text{and} \Ket{q_{l+1}}=\Ket{1} \\
		0 & \mbox{if } \Ket{q_l}=\Ket{1} \text{and} \Ket{q_{l+1}}=\Ket{0}
	\end{array}
\right.
\end{split}
\end{equation}

\begin{equation}
\label{eqn:rot_ql1}
\begin{split}
&\theta_{l+1, q_l=\Ket{1}}  = 2\times \tan^{-1} \sqrt{\dfrac{P(q_{l+1}=\Ket{1}|q_l=\Ket{1})}{P(q_{l+1}=\Ket{0}|q_l=\Ket{1})}}\\
 & = 2\times \tan^{-1}\sqrt{\dfrac{P(q_{l+1}=\Ket{1},q_l=\Ket{1})}{P(q_{l+1}=\Ket{0},q_l=\Ket{1})}} = 2\times \\
& \tan^{-1}\sqrt{\dfrac{\sum_{j=0}^{2^{m_i}-1} \alpha_j^2 \mathbb{I}_{q_l=\Ket{1}, q_{l+1}}(\Ket{q_l q_{l+1} \dots q_{l+m_i-1}}_j)}{1-\sum_{i=0}^{2^{m_i}-1} \alpha_j^2 \mathbb{I}_{q_l=1, q_{l+1}}(\Ket{q_l q_{l+1} \dots q_{l+m_i-1}}_j)}}
\end{split}
\end{equation}

\noindent$CCU_{2 \dots m_i-1, q_l=\Ket{1}, q_{l+1}=\Ket{1}}$ and $CCU_{2 \dots m_i-1, q_l=\Ket{1}, q_{l+1}=\Ket{0}}$ are again decomposed into a set of controlled-controlled single qubit rotations and triply controlled $m_i$-3 qubit rotations following the procedure described above. This decomposition is carried out until we reach $m_i-1$ controlled qubit rotations implemented on the qubit $q_{l+m_i-1}$. In this way, a $m_i$-qubit rotation required to realize the probabilities associated with a discrete variable with more than two states is achieved through uncontrolled/controlled/multi-controlled qubit rotations. Multi-controlled qubit rotations can be implemented using ancilla qubits as detailed in Figure \ref{fig:decompose_CnRY}. When the multi-state variable is a child node, then we will have a different $m_i$ qubit rotation for each combination of the parent nodes. Depending on the number of parent nodes, Each controlled/multi-qubit controlled $m_i$-qubit rotation can be represented following the above sequential decomposition process. 

In Section \ref{subsec:2node_child}, we discussed the implementation of controlled rotations to realize conditional probabilities of a child node when both the parent and child nodes have two states. Here, let us consider the cases when a combination of multi-state variables and two-state variables are parent nodes for a multi-state child node. Consider a variable $V_i$ with $n_i$ states with $\Pi_{V_i}$ as the set of parent nodes. Let $\Pi_{V_{i,j}}$ represent the $j^{th}$ parent node and $n_{\Pi_{V_{i,j}}}$ represent the number of discrete states in the $j^{th}$ parent node. Number of qubits required to represent $\Pi_{V_i}$ can be calculated as

\begin{equation}
\label{eqn:nq_parent}
 m_{q, \Pi_{V_i}} = \sum_{i=1}^{|\Pi_{V_i}|} \ceil{\log_2 n_{\Pi_{V_{i,j}}}}
\end{equation}

\noindent where $m_{q, \Pi_{V_i}}$ is the number of qubits required to represent $\Pi_{V_i}$ and $|\Pi_{V_i}|$ represents the cardinality of the set of parent nodes. If $n_i$ represents the number of states of child node $V_i$, then the highest order of $C^nR_Y$ gate required to realize the conditional probabilities of $V_i$ can be calculated as $n = m_{q, \Pi_{V_i}} + \ceil{\log_2 n_i}$. In order to implement this $C^nR_Y$ gate, we will need $n-1 = m_{q, \Pi_{V_i}} + \ceil{\log_2 n_i}-1$ ancilla qubits. Therefore, the total number of qubits required to obtain a Bayesian network with a combination of two-state and multi-state variables is given as

\begin{equation}
\label{eqn:nBN}
m_{BN} = \Bigg(\sum_{i=1}^m \ceil{\log_2 n_i}\Bigg) + \max_i \Big(m_{q, \Pi_{V_i}} + \ceil{\log_2 n_i} - 1\Big)    
\end{equation}

\noindent where $m_{BN}$ denoted the number of qubits required to represent a given BN, $m_{q, \Pi_{V_i}} + \ceil{\log_2 n_i} - 1$ is the number of ancilla qubits required to realize the conditional probabilities of node $V_i$. As mentioned in Section \ref{subsec:2node_child}, the same set of ancilla qubits can be used for various child nodes. Since qubits can represent only discrete states, any continuous variables need to be discretized to be represented using qubits. If a continuous variable is discretized into more than two states, then the above procedure to handle discrete variables with more than two levels can be used. If the discretization involves only two states, a single qubit can be used to represent it. 


\section{Illustration Examples}
\label{sec:examples}

For illustration of the proposed methodology, we consider three examples with varying properties from the financial industry: (1) a 4-node Bayesian network for an oil company stock price prediction; (2) a 10-node Bayesian network used for liquidity risk assessment; and (3) a Naive Bayes classifier with 8 features (a total of 9 nodes) used for bankruptcy prediction.

The goal of the examples is to demonstrate the application of proposed methods to represent generic discrete quantum Bayesian networks on a gate-based quantum computing platform. The proposed approach is generic and can be used to represent any discrete Bayesian network with any number of nodes. Representation of larger Bayesian network will result in quantum circuits with large number of gates but the procedure to construct the quantum circuits remain the same for smaller or larger Bayesian networks.

The first example is a simple 4-node Bayesian network with binary variables; this example was used to demonstrate the fundamental approach for the representation of unconditional and conditional probabilities using single-qubit rotation, controlled rotations, and use of an ancilla qubit to represent conditional probabilities with two parent nodes. 

The second example is a slightly larger Bayesian network with ten binary variables, where one of the variables has three parent nodes. This example was used to demonstrate that the proposed approach can be used to demonstrate the scalability of the proposed framework, and also the use of two ancilla qubits to realize higher-order controlled rotations.

The third example is used to demonstrate the representation of Bayesian networks with nodes having more than two states using qubits, which have only two states. The goal of this example is to demonstrate that the proposed framework can be used to represent any generic discrete Bayesian network with nodes having more than two states using qubits.

For each of the three examples, we design quantum circuits using the proposed C-QBN approach. We performed the quantum computing simulations using a Python package called Quantum Information Science kit (Qiskit), which uses Open QASM or Open Quantum Assembly Language developed by IBM, and is used in all the IBM quantum hardware \mbox{\citep{IBMQE}}. The histograms and box plots are built using the matplotlib library \mbox{\citep{hunter2007matplotlib}}.

We demonstrate the proposed methods on a simulated platform instead of using real quantum computers as hardware implementation of circuits of large depths (large number of gates) are affected by noise, which leads to incorrect results \citep{mandviwalla2018implementing, martin2019towards}. Since simulations are not affected by hardware noise, we use them to demonstrate the proposed methods. 

The probabilities of various states of the BNs are computed, and these results are compared with the probabilities obtained from simulating the examples on a classical Bayesian network platform such as Netica \citep{Netica}. 

Through this paper, we demonstrate the representation of marginal and conditional probabilities of discrete random variables using elementary single-qubit and two-qubit quantum gates. Moreover, we develop a systematic approach to represent nodes with more than two states as previous literature only focused on nodes with two states; this facilitates the representation of any generic discrete Bayesian network.

\subsection{4-node BN: Oil Company Stock Price}
\label{subsec:4nodeBN}

\begin{center}
\begin{figure}
\includegraphics[width=0.95\columnwidth]{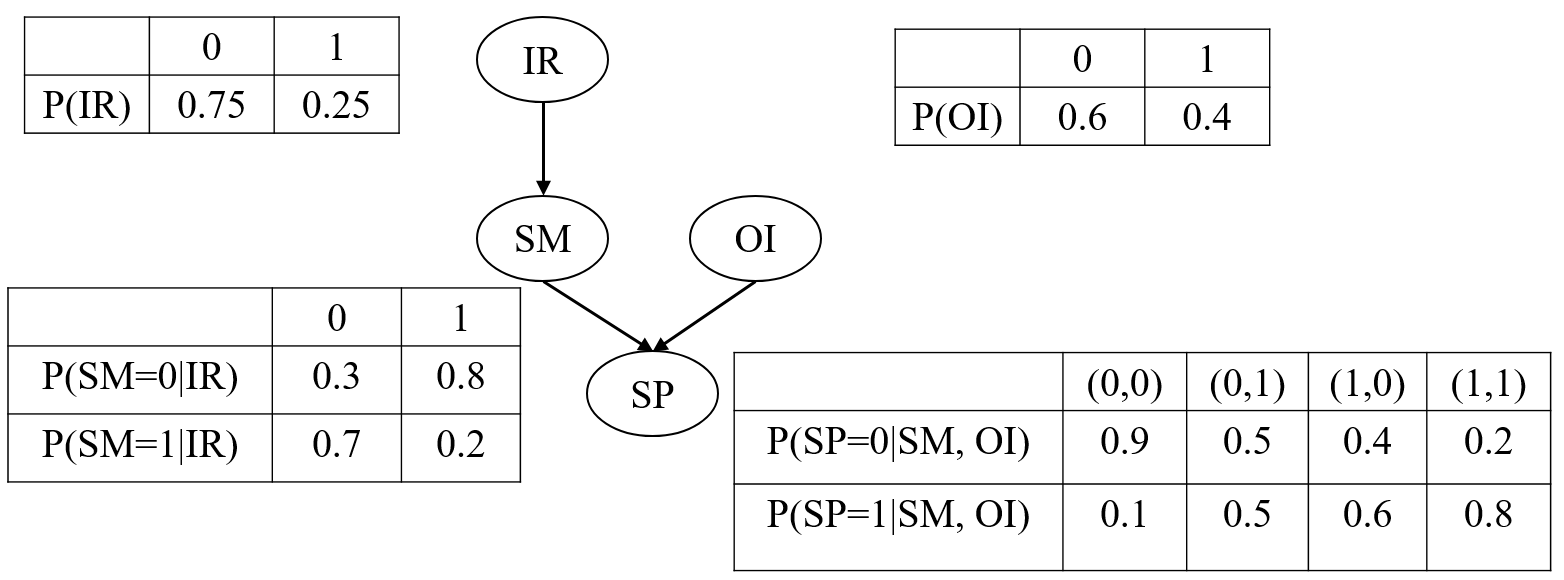}
\caption{A 4-node Bayesian network for an oil company stock price prediction} \citep{shenoy2000bayesian}
\label{fig: 4nodeBN}
\end{figure}
\end{center}

This 4-node Bayesian network example to assess an oil company stock price is obtained from \citep{shenoy2000bayesian}. The four variables in this network are the interest rate (IR), stock market (SM), oil industry (OI), and oil company stock price (SP). IR has two states - \textit{high} and \textit{low}; SM has two states - \textit{good} and \textit{bad}; OI has two states - \textit{good} and \textit{bad}; and SP has two states - \textit{high} and \textit{low}. Here, we represent \textit{low/bad} with state 0 and \textit{high/good} with state 1. The dependence between these four variables and associated conditional probability tables are given in Figure \ref{fig: 4nodeBN}.

The BN in Figure \ref{fig: 4nodeBN} has two root nodes, i.e. nodes without parent nodes (IR, OI), one node with only one parent node (SM), and one node with two parent nodes (SP). Since SP has two parent nodes, we use one ancilla qubit to represent its conditional probability values as discussed in Section \ref{subsec:2node_child}. This results in a five-qubit system (four qubits to represent four variables in Figure \ref{fig: 4nodeBN} and an ancilla qubit). We discuss below the construction of quantum circuits corresponding to this BN using the C-QBN approach described in Section \ref{sec:method}. 

\begin{figure}[H]
\centering
\includegraphics[scale=0.3]{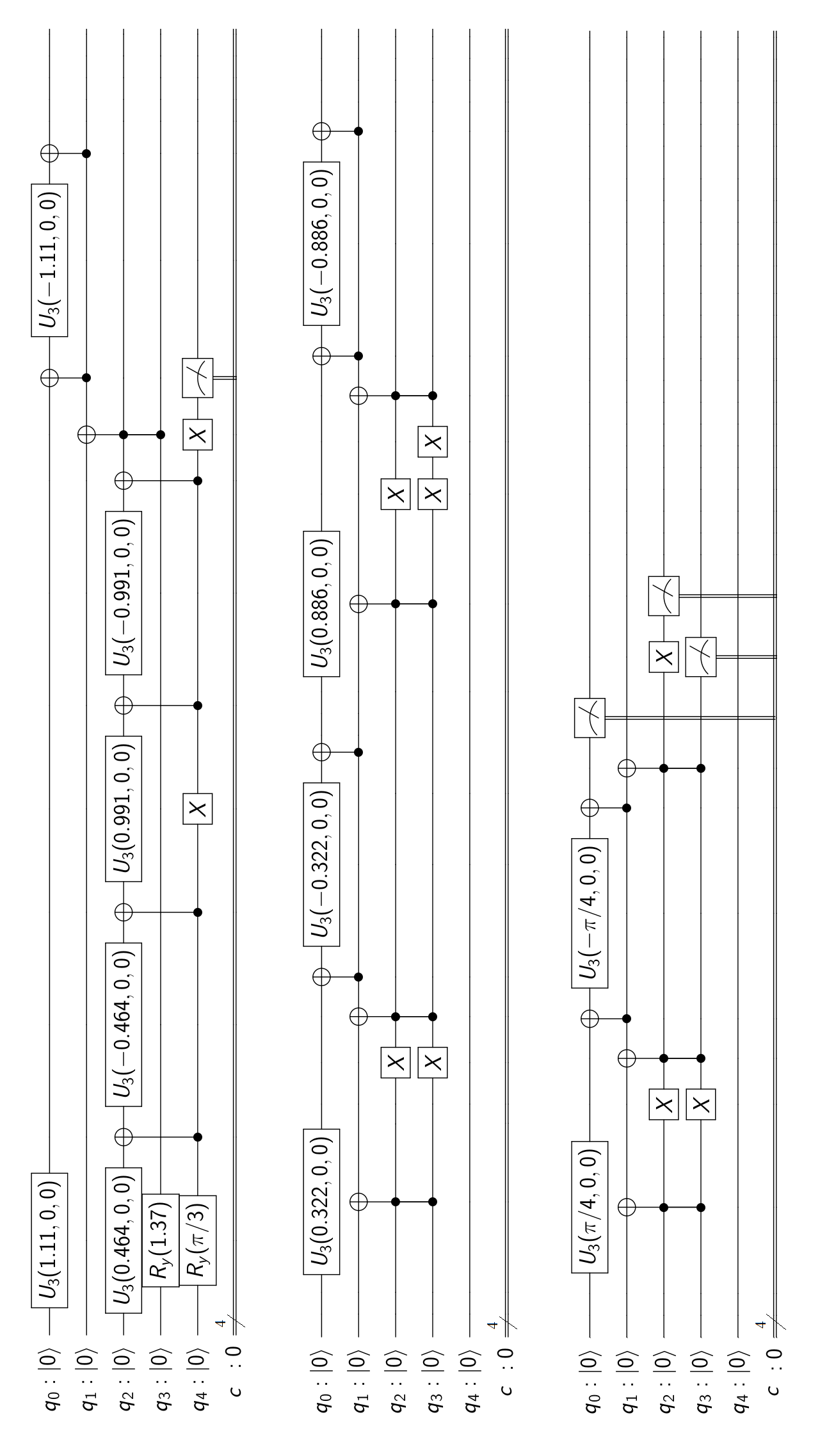}
\caption{Quantum circuit of the 4-node oil company stock price BN. Variables IR, OI, SM and SP are mapped to $q_4$, $q_3$, $q_2$, and $q_0$ respectively, $q_1$ is the ancilla qubit, and $c$ represents the classical bits used to store of values of qubits after measurement} 
\label{fig: 4qbnlatex_cqbn}
\end{figure}

\textbf{Quantum circuit:} Figure \ref{fig: 4qbnlatex_cqbn} provides the quantum circuit corresponding to the BN in Figure \ref{fig: 4nodeBN} constructed using the C-QBN approach. The five qubits are denoted as $q_i, i=0\dots 4$ and the measurement bit is denoted as $c$.

The variables - IR, OI, SM, and SP are denoted using the qubits $q_4, q_3, q_2$ and $q_0$ respectively, and the ancilla qubit is $q_1$. We chose $q_1$ as the ancilla qubit for the purpose of illustration. In reality, any qubit can be chosen as an ancilla qubit in Qiskit. We used this mapping as the representation of an n+1-qubit state is given as $\Ket{q_nq_{n-1}\dots q_0}$, i.e., the state of the $n+1^{th}$ qubit ($q_n$) is written first while the first qubit ($q_0$) is written at the end. By following this mapping, the parent nodes appear ahead of their associated child nodes.

After mapping the variables to various qubits, we now identify the appropriate gates to be implemented on those qubits to obtain the required marginal or conditional probability values. Let us begin with the root nodes (IR and OI). The rotation angles required to represent those root nodes were calculated using Eq. (\ref{eqn:theta}) as $\theta_{IR} = 2\times \tan^{-1}\bigg(\sqrt{\dfrac{0.25}{0.75}}\bigg) = \dfrac{\pi}{3}$ and $\theta_{OI} = 2\times \tan^{-1}\bigg(\sqrt{\dfrac{0.4}{0.6}}\bigg) = 1.37$.
 For realizing child nodes, we compute the associated rotation angles for various combinations of the parent node(s).
 
 To represent SM node, we compute its rotation angles when IR=0 and IR=1 as $\theta_{SM,0} = 2\times \tan^{-1}\bigg(\sqrt{\dfrac{0.7}{0.3}}\bigg) = 1.982$ and $\theta_{SM,1} = 2\times \tan^{-1}\bigg(\sqrt{\dfrac{0.2}{0.8}}\bigg)=0.928$ respectively. Here, $\theta_{SM,j}$ corresponds to the rotation angle of SM for a given value $j$ of its parent node, IR. As discussed in Section \ref{subsec:2node_child}, the controlled rotations are decomposed into a combination of uncontrolled rotations and CNOT gates. For example, the conditional probability values of SM when IR=1 are realized by implementing a controlled-rotation gate $CR_Y(\theta_{SM,1})$. Following Section \ref{subsec:2node_child}, this gate is implemented as $\bigg(I\otimes R_Y\bigg(\dfrac{\theta_{SM,1}}{2}\bigg)\bigg)CX\bigg(I\otimes R_Y\bigg(\dfrac{-\theta_{SM,1}}{2}\bigg)\bigg)CX$. The controlled-rotation when IR=0 is implemented by first flipping the states of the $q_4$ (IR) qubit using an $X$ gate, and then applying the $CR_Y(\theta_{SM,0})$ gate. Thus, the conditional probability values of SM are realized for various values of its parent node (IR). We now consider the SP node.

Since SP has two parent nodes, we have four rotation angles ($\theta_{SP,00}$, $\theta_{SP,01}$, $\theta_{SP,10}$, $\theta_{SP,11}$) for various values of the two parent nodes; these values were calculated using Eq. (\ref{eqn:theta}) as $0.644, 1.772, \frac{\pi}{2}, 2.22$ respectively. Here, $\theta_{SP,jk}$ corresponds to the rotation angle of SP for given values $j$ and $k$ of parent nodes OI and SM respectively. Following Section \ref{subsec:multistate}, the controlled-controlled rotations are implemented using an ancilla qubit. At the end of the circuit, we add the measurement gates and stores the measured qubit values in a classical bit register ($c$ in Figure \ref{fig: 4qbnlatex_cqbn}). We consider only four measurements across the qubits associated with various nodes in the BN, and do not consider a measurement gate for the ancilla qubit as it does not represent any variable in the BN.

\textbf{Circuit simulation:} The BN is simulated using the circuit constructed with the C-QBN approach, and the accuracy of the results is compared with those obtained using Netica, which is a classical Bayesian network software. 
After a simulation is made, the system is measured, which returns a single quantum state (such as $\Ket{1010}$). A total of 8192 shots were carried out in each simulation, and the measured states after each shot are used to estimate the the probabilities of all the states. We used 8192 shots as that was the maximum number of shots possible on the real IBM quantum computers such as the 5-qubit IBM QX5 \citep{mandviwalla2018implementing}.

When a measurement is made, one of the 16 states is observed as we were measuring only four qubits corresponding to four nodes in the BN. The probability associated with each state ($P(\Ket{q_4q_3q_2q_0})$) can be computed using the Monte Carlo approach as 
\begin{equation}
P(\ket{q_4q_3q_2q_0}) = \dfrac{n_{\ket{q_4q_3q_2q_0}}}{N}
\end{equation}

\noindent where $P(\Ket{q_4q_3q_2q_0})$ is the probability of state $\Ket{q_4q_3q_2q_0}$; $n_{\Ket{q_4q_3q_2q_0}}$ and $N$ represent the number of times $\Ket{q_4q_3q_2q_0}$ is observed and the total number of shots  (8192) respectively. The marginal probabilities of each of the nodes can be estimated using Equation by marginalizing over the state probabilities calculated using Eq. (\ref{eqn:marginalize}).

\begin{equation}
\label{eqn:marginalize}
P(\Ket{q_i}) = \sum_{q_j, j=4,3,2,0, j\neq i} P(\Ket{q_4q_3q_2q_0})
\end{equation}


\textbf{Comparison of simulation results:} As discussed above, we ran 8192 shots of the quantum circuit to estimate the marginal probabilities. Since the marginal probabilities are estimated from data, there could be variation across multiple runs of the quantum circuit. In order to quantify the variation across runs, we ran the circuit $r$ times and obtained the marginal probability values from each run. Given the simulation results from $r$ runs, we compute the $(1-\alpha)$ confidence intervals of the estimated marginal probabilities and checked if the marginal probabilities from Netica fall within the estimated intervals. 

Let $p_i^m, i=1\dots r, m={IR, OI, SM, SP}$ represent the marginal probability value of the $m^{th}$ variable in the $i^{th}$ run, then the sample mean and standard deviation can be calculated as $\bar{p}^m = \dfrac{\sum_{i=1}^{r} p_i^m}{r}$ and $s^m = \sqrt{\dfrac{\sum_{i=1}^{r} (p_i^m-\bar{p}^m)^2}{r-1}} $ respectively. Given the sample mean and standard deviation, the $(1-\alpha)$ confidence interval were calculated as $\bar{p}^m \pm t_{\frac{\alpha}{2}} \dfrac{s^m}{\sqrt{r}}$. Here, $t_{\frac{\alpha}{2}}$ is the t-statistic corresponding to the $(1-\alpha)$ confidence interval.  In this study, we chose $r=10$ and $\alpha=0.05$. The sample mean and standard deviation, and the 95\% CIs of the marginal probabilities using both the methods are provided in Table \ref{tab:4node_marginals}, along with the marginal probabilities obtained using Netica. The variation in the probability values obtained using the C-QBN approach, can be attributed to the variability in the measurement process. 

\begin{center}
\begin{figure}[H]
\includegraphics[width=0.98\columnwidth]{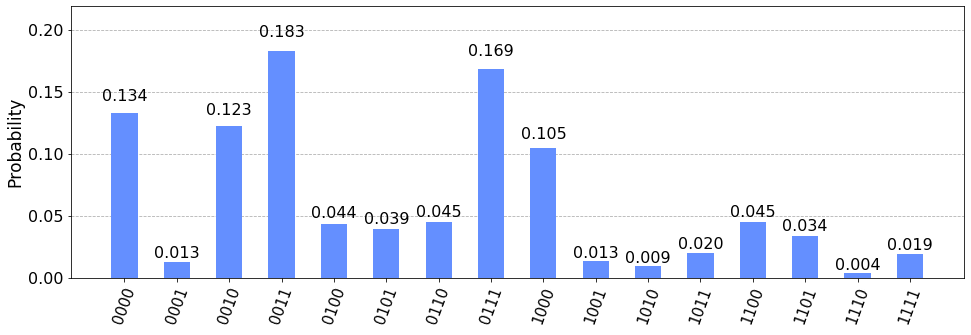}
\caption{Simulation results of the oil company stock price QBN using Qiskit with 8192 shots (quantum states on the horizontal axis and  joint probabilities on the vertical axis)}
\label{fig: 4qb_histogram_qiskit}
\end{figure}
\end{center}

\begin{table}
\caption{Comparison of marginal probabilities in the 4-node Bayesian network with Netica (classical computation) and the proposed approach (C-QBN)}
\label{tab:4node_marginals}
\setlength{\arrayrulewidth}{0.2mm}
\setlength{\tabcolsep}{7pt}
\renewcommand{\arraystretch}{1}

\begin{center}
\begin{tabular}{ |c|c|c|c|c|}

\hline
& \textbf{Netica} & \multicolumn{3}{c|}{\textbf{C-QBN}} \\
\hline
Value &  Probability   & Mean & SD & 95\% CI \\
\hline
IR=0 &  0.750  & 0.750 &  0.008 &[0.747, 0.754] \\
\hline
SM=0 &  0.425   & 0.425 & 	0.006 &[0.422, 0.428] \\
\hline
OI=0 &  0.600 & 0.601	& 0.003 &[0.598, 0.604]\\
\hline
SP=0   & 0.499 &  0.499 &  0.004 &[0.495, 0.502]\\
\hline
\end{tabular} 
\end{center}
\end{table}

From Table \ref{tab:4node_marginals}, it can be observed that the marginal probabilities from Netica fall within their estimated confidence intervals obtained from the quantum circuit simulations. Since each node has two states, we provided the probabilities of only one of the states as the probabilities of the other states can be computed from the given states. For example, $P(IR=1)=1-P(IR=0)$. Thus, the C-QBN approach was used to represent the 4-node Bayesian network.

\subsection{10-node BN: Liquidity Risk Assessment}
\label{subsec:liquid}

\begin{figure}[htb]
\centering
	\includegraphics[scale=0.3]{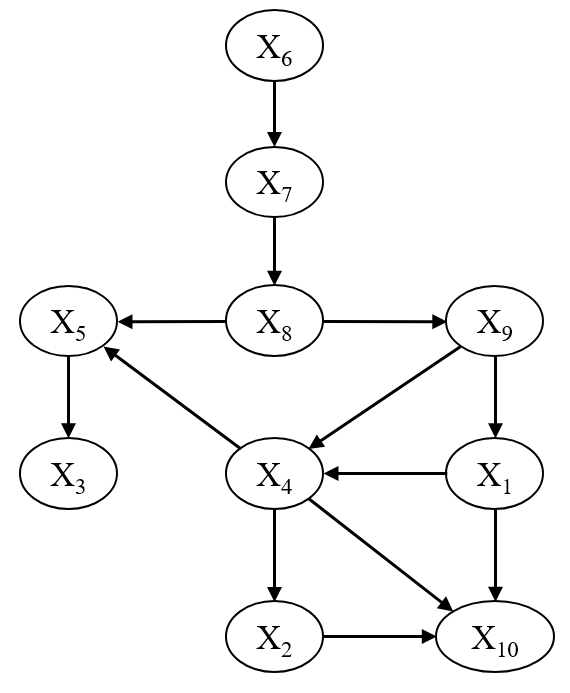}
	\caption{A 10-node Bayesian network for liquidity risk assessment\citep{tavana2018artificial}}
\label{fig:liquid}
\end{figure}

\begin{figure}[htb]
\begin{center}
\includegraphics[width=0.98\columnwidth]{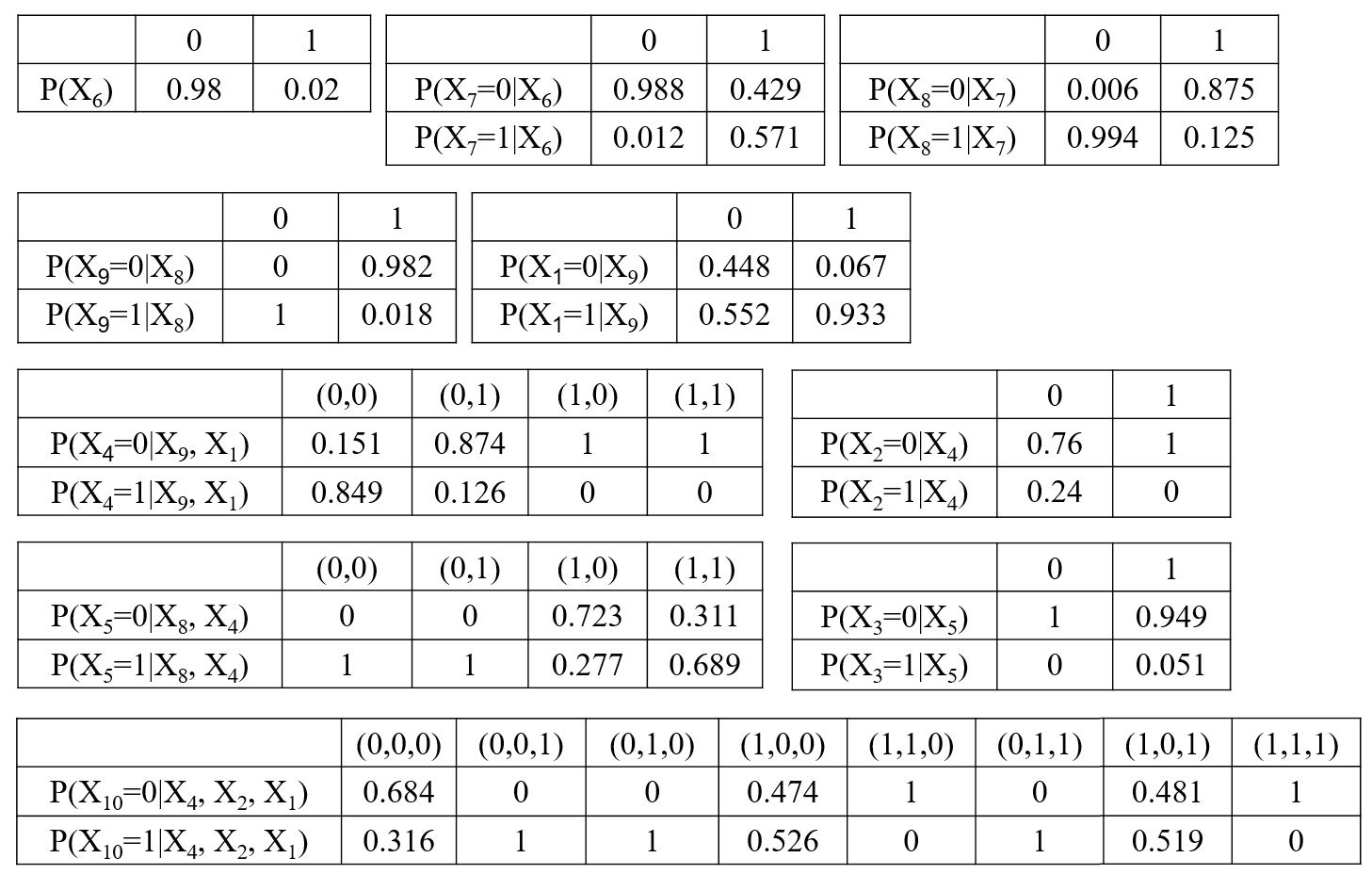}
\caption{Marginal and conditional probabilities of various nodes in the liquidity risk assessment Bayesian network \citep{tavana2018artificial}}
\label{fig:ex2_cpt}
\end{center}
\end{figure}

Here, we consider designing a quantum circuit to represent a 10-node Bayesian network obtained from \citep{tavana2018artificial} used for liquidity risk assessment in banking. The 10 variables in the Bayesian network are described in Table \ref{tab:10BN}, and the dependence between various variables are shown in Figure \ref{fig:liquid}, and the conditional probability tables are given in Figure \ref{fig:ex2_cpt}.

 In Table \ref{tab:10BN}, \textbf{B} refers to the bank under consideration, and \textbf{O} refers to other banks. This BN has one root node ($X_6$), six nodes with one parent node ($X_7, X_8, X_9, X_1, X_2, X_3$), two nodes with two parent nodes ($X_4, X_5$), and one node with three parent nodes ($X_{10}$). Since the maximum number of parent nodes is three, we need two ancilla qubits to represent the conditional probability values in addition to the ten qubits used to represent the ten nodes in the BN totaling to 12 qubits. The representation of the root node ($X_6$), child nodes with either one or two parent nodes follows the same procedure as detailed in Section \ref{subsec:4nodeBN}. Therefore, we discuss below the representation of $X_{10}$, which is the child node with three parent nodes ($X_1, X_2, X_4$) using the C-QBN approach.

\begin{table}[H]
\centering
\caption{Variables in the 10-node liquidity risk assessment Bayesian network \citep{tavana2018artificial}}
\label{tab:10BN}
\setlength{\arrayrulewidth}{0.2mm}
\setlength{\tabcolsep}{3pt}
\renewcommand{\arraystretch}{1.5}
\begin{tabular}{|p{2cm}|p{8cm}|}
 \hline
 \textbf{Variable}  & \textbf{ Description} \\[1ex]
 \hline
 $X_1$ & $\text{Liquidity ratio} = \dfrac{\text{Liquid assets of \textbf{B}}}{\text{Current liabilities of \textbf{B}}}$\\[1ex]
  \hline
 $X_2$ & $\dfrac{\text{Credits of \textbf{B} in \textbf{O}}}{\text{Credits of \textbf{O} in \textbf{B}}}$\\[1ex]
 \hline
 $X_3$ & $\dfrac{\text{Long term deposits of \textbf{B}}}{\text{Short term deposits of \textbf{B}}}$\\[1ex]
 \hline
 $X_4$ & $\dfrac{\text{Credits of \textbf{B} in \textbf{O}}}{\text{Credits of \textbf{O} in \textbf{B}}}$\\[1ex]
 \hline
 $X_5$ & $\dfrac{\text{Total loan of \textbf{B}}}{\text{Total deposits of \textbf{B}}}$\\[1ex]
 \hline
  $X_6$ & $\dfrac{\text{bonds of \textbf{B}}}{\text{Total assets of \textbf{B}}}$\\[1ex]
  \hline
  $X_7$ & $\dfrac{\text{Volatile deposits of \textbf{B}}}{\text{Total liabilities of \textbf{B}}}$\\[1ex]
 \hline
   $X_8$ & $\dfrac{\text{Short investments of \textbf{B}}}{\text{Total assets of \textbf{B}}}$\\[1ex]
\hline
$X_9$ & $\dfrac{\text{Credits of \textbf{B} in central bank}}{\text{Total deposits of \textbf{B}}}$\\[1ex]
\hline
$X_{10}$ & $\text{Bank liquidity risk} = \dfrac{\text{Long term deposits of \textbf{B}}}{\text{Short term deposits of \textbf{B}}}$\\[1ex]
\hline
\end{tabular}
\end{table}

\textbf{Quantum circuit:} Figure \ref{fig:10nodeBN_cqbn} provides the quantum circuit of the 10-node BN constructed the C-QBN approach. The 12 qubits are denoted as $q_i, i=0\dots 11$ and the measurements of various qubits are stored in classical bits denoted as $c$. In this circuit $X_1$ is mapped to qubit $q_7$, $X_2$ represented by $q_3$, $X_3$ to $q_4$, $X_4$ to $q_6$, $X_5$ to $q_5$, $X_6$ to $q_{11}$, $X_7$ to $q_{10}$, $X_8$ to $q_9$, $X_9$ to $q_8$. Finally, $X_{10}$ is represented using $q_0$, and $q_1, q_2$ are the ancilla qubits. Since $X_{10}$ are three parent nodes, we need to implement $C^3R_Y$ gate for each of the 8(=$2^3$) combinations of the parent nodes. Following Section \ref{subsec:2node_child}, the $C^3R_Y$ gate requires the use of two ancilla qubits, and is build through a combination of one, two, and three-qubit gates($R_Y$, CNOT, CCNOT).

For example, the conditional probability values of $X_5$ when $X_8=1$ and $X_4=0$ would be obtained by implementing a controlled-controlled rotation gate $CCR_Y$. Since $X_5$ has two parent nodes, we have four rotation angles($\theta_{X_5,00}$, $\theta_{X_5,01}$, $\theta_{X_5,10}$, $\theta_{X_5,11}$) for various values of the two parent nodes. Following Section \ref{subsec:multistate}, the controlled-controlled rotations for $X_5$ are implemented using one of the two available ancilla qubits ($q_2$ is used here).

\textbf{Circuit simulation and results}: Following the 4-node BN, we ran the quantum circuits obtained using the C-QBN ten times each with 8192 shots. The marginal probabilities of nodes from Netica, along with the sample mean, sample standard deviation and the 95\% confidence intervals of the sample mean are provided in Table \ref{tab:10node_comparison}.

\begin{center}
\begin{figure}[H]
\includegraphics[width=0.99\columnwidth]{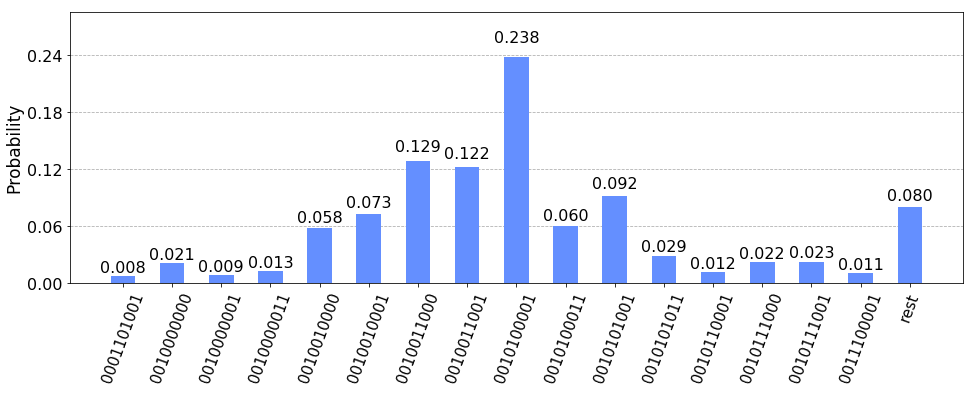}
\caption{Simulation results of the liquidity risk assessment QBN using Qiskit with 8192 shots (quantum states on the horizontal axis and  joint probabilities on the vertical axis)}
\label{fig: 4qb_sim_ibm}
\end{figure}
\end{center}

\begin{table}
\begin{center}
\setlength{\arrayrulewidth}{0.2mm}
\setlength{\tabcolsep}{7pt}
\renewcommand{\arraystretch}{1}
\caption{Comparison of marginal probabilities in the 10-node Bayesian network with Netica and C-QBN}
\label{tab:10node_comparison}
\begin{tabular}{|c|c|c|c|c|}
\hline

& \textbf{Netica} & \multicolumn{3}{c|}{\textbf{C-QBN}} \\
\hline
Value & Probability & Mean & SD & 95\% CI  \\
\hline
$X_1=0$ &  0.431 & 0.432 & 0.007 & [0.429,  0.433]\\
\hline
$X_2=0$ &  0.863   & 0.864 	&	0.003 &[0.862, 0.865]\\
\hline
$X_3=0$ & 0.976 & 0.976      &	0.002 &[0.975, 0.976]\\
\hline
$X_4=0$   & 0.570 &  0.570 	&	0.006&[0.568,	0.571]\\
\hline
$X_5=0$  & 0.527  &   0.525	&	0.004&[0.524,	0.529]\\
\hline
$X_6=0$ &  0.980 &   0.980  	&	0.001&[0.980,	0.981]\\
\hline
$X_7=0$ & 0.977  &   0.976     &	0.001&[0.976,	0.977]\\
\hline
$X_8=0$ & 0.026  &  0.027      &	0.001&[0.026,	0.027]\\
\hline
$X_9=0$ &  0.956 &  0.956 	 &	0.002&[0.955,	0.957]\\
\hline
$X_{10}=0$ & 0.240  &    0.239	&	0.006&[0.238, 0.241]\\
\hline
\end{tabular}
\end{center}
\end{table}
 From the results in Table \ref{tab:10node_comparison}, it can be observed that the true marginal probabilities (obtained from Netica) fall within the confidence intervals obtained using the C-QBN approach. These results help conclude that the C-QBN approach is able to simulate the 10-node Bayesian network with two and three parent nodes, each with two states.

\subsection{9-node Naive Bayes Classifier: Bankruptcy Prediction}

Here, we consider designing a quantum circuit to represent a Naive Bayes classifier used for bankruptcy prediction; this model is obtained from \citep{sun2007using}. There are eight features in this classifier that correspond to several financial-accounting, market-based, and other extraneous factors. The financial-accounting factors are Cash/Total Assets (CH), a variable related to the variation in cash and short-term marketable securities (LM), a binary variable to check if the net income was negative in the last two years (IT), and a variable that represents the ratio of the change in net income and the sum of the absolute net income in the last two years (CHN). The market-based factors are a variable that represents the natural logarithm of firm's size relative to the CRSP NYSE/AMEX/NASDAQ market capitalization index (M) and a variable that represents the difference of the firm's stock return and the value-weighted CRSP NYSE/AMEX/NASDAQ index return in the previous year (R). 

\begin{figure}[H]
\begin{center}
\includegraphics[width=0.98\columnwidth]{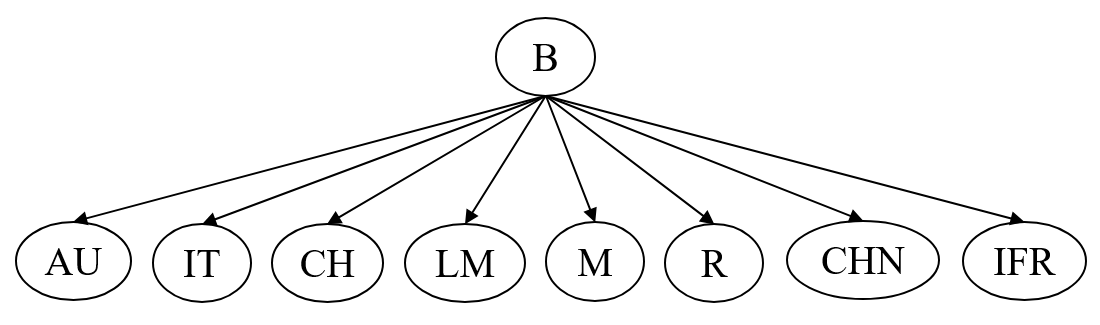}
\caption{A 9-node Naive Bayes classifier for bankruptcy prediction \citep{sun2007using}}
\label{fig: naivebayes}
\end{center}
\end{figure}

\begin{figure}[htb]
\begin{center}
\includegraphics[width=0.98\columnwidth]{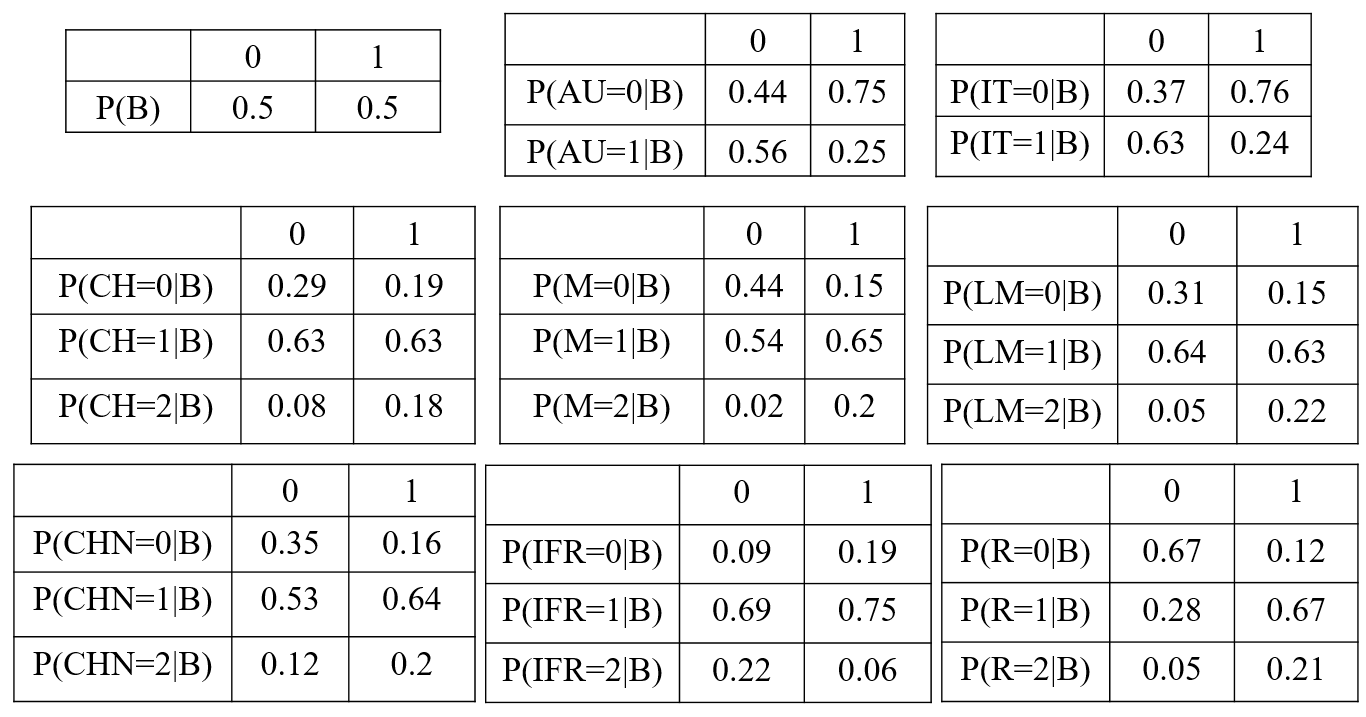}
\caption{Marginal and conditional probabilities of various nodes in the bankruptcy prediction naive Bayes classifier \citep{sun2007using}}
\label{fig: naive_CPT}
\end{center}
\end{figure}

The extraneous factors are variables that relate to the Compustat codings (AU) and Industry Failure Rate (IFR). The bankruptcy classification status is denoted with the variable B. Figure \ref{fig: naivebayes} shows the Naive Bayes model, and the associated conditional probability tables are provided in Figure \ref{fig: naive_CPT}. The variables B, AU and IT have two states $\{0,1\}$ while all the remaining variables have three states $\{0,1,2\}$. We are using this example for the sake of illustration, and the readers are referred to \citep{sun2007using} for more details about the variables and the model. Since each the variables B, AU and IT has two states, it can be represented using a single qubit. Each of the remaining variables has three states; therefore, each node is represented using $\ceil{\log_2 3} = 2$ qubits. 
The total number of qubits used to represent the 9-node Naive Bayes classifier is 16 ($3\times1 + 6\times2 + 1$). The representation of the root node (B), and child nodes with one parent node (AU, IT) follows the procedure described in Section \ref{subsec:4nodeBN}. Here, we discuss the representation of child nodes with more than two levels (CH, LM, M, R, CHN, IFR) using the C-QBN approach.

\begin{figure}
\centering
\includegraphics[scale=0.225]{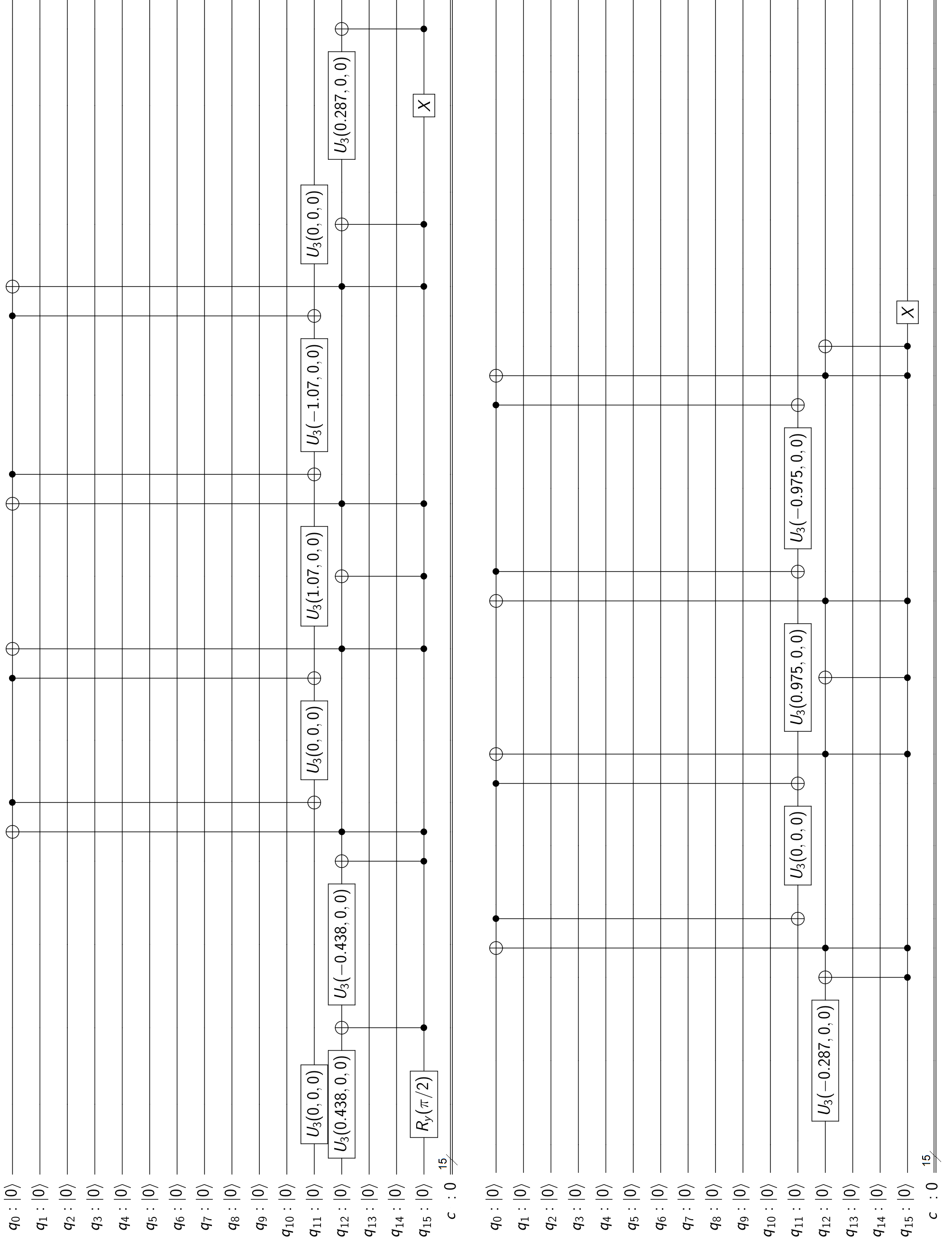}
\caption{Quantum circuit representing nodes B and CH in the 9-node naive Bayes classifier. Variables B, AU, IT, CH, LM, M, R, CHN, and IFR are mapped to $q_{15}, q_{14},q_{13}, (q_{12}, q_{11}),(q_{10}, q_{9}),(q_{8}, q_{7}),(q_{6}, q_{5}), (q_{4}, q_{3}),$ and $(q_{2}, q_{1})$ respectively, $q_0$ is the ancilla qubit, and $c$ represents  classical bits.} 
\label{fig: B-CH}
\end{figure}

\textbf{Quantum circuit:}  The 16 qubits are denoted as $q_i, i=0 \dots 15$ and the measurements of various qubits are stored in the classical bit register $c$. The nine variables B, AU, IT, CH, LM, M, R, CHN, and IFR are mapped to $q_{15}, q_{14},q_{13}, (q_{12}, q_{11}),(q_{10}, q_{9}),(q_{8}, q_{7}),(q_{6}, q_{5}), (q_{4}, q_{3}),$ and $(q_{2}, q_{1})$ respectively. Any qubit can be chosen as the ancilla qubit, and in this example, we chose $q_0$ for the sake of illustration. We discuss the representation of CH, and the same procedure can be applied to other nodes as well. We map the three states of CH \{0,1,2\} to the states $\ket{00}$, $\ket{01}$, $\ket{10}$ of qubits $q_{12}$ and $q_{11}$. Figure \ref{fig: B-CH} shows the associated quantum circuit with gates associated with B and CH nodes only. We apply the appropriate $CU$ transformations to realize the conditional probability values for various values of B \{0,1\}. Here, the control qubit is $q_{15}$ and the target is a two-qubit system $(q_{12}, q_{11})$. 

First, let us consider the case when $q_{15}=\Ket{1}$, i.e., B=1. When $q_{15}=\Ket{1}$, the transformation $U$ should result in the probability values of 0.19, 0.63 and 0.18 for $(q_{12}, q_{11})$ states of $\ket{00}$, $\ket{01}$ and $\ket{10}$ respectively. The probability of state $\Ket{11}$ is fixed at 0. The multi-qubit rotation $U$ is not an elementary transformation, we will decompose it into a combination of one and two-qubit elementary transformations. The probability of $\Ket{0}$ and $\Ket{1}$ states of $q_{12}$ can be calculated as $P(\Ket{00}) + P(\Ket{01}) = 0.19+0.63 = 0.82$ and $P(\Ket{10}) + P(\Ket{11}) = 0.18 + 0 = 0.18$ respectively. We realize the marginal probabilities of $q_{12}$ using a single-qubit $R_Y$ gate with the rotation angle $\theta_{q_{12}, q_{15} = \Ket{1}} = 2\times \tan^{-1}\bigg(\sqrt{\dfrac{0.18}{0.82}}\bigg) = 0.5725$. 
 
After realizing the marginal probabilities of $q_{12}$, we consider the conditional probabilities of $q_{11}$ given $q_{12}$. When $q_{12}=\Ket{1}$, the probability of $q_{11}=\Ket{0}$ is computed using Eq. (\ref{eqn:cond_prob}) as

\begin{equation}
\label{eqn:cond_prob}
\begin{split}
P(q_{11}=\Ket{0}|q_{12}=\Ket{1}) &= \dfrac{P(q_{12}=\Ket{1},q_{11}=\Ket{0})}{P(q_{12}=\Ket{1})}\\
&= \dfrac{0.18}{0.18} = 1 
\end{split}
\end{equation}

Therefore, $P(q_{11}=\Ket{1}|q_{12}=\Ket{1}) = 1-P(q_{11}=\Ket{0}|q_{12}=\Ket{1}) = 0$. Thus, the probability values associated with the $\ket{10}$ and $\ket{11}$ states are applied through a $CR_Y$ where $q_{12}$  and $q_{11}$ are the control and target qubits respectively, and with a rotation angle, $\theta_{q_{11}, q_{12}=\Ket{1} , q_{15}=\Ket{1}} = 2\times \tan^{-1}\bigg(\sqrt{\dfrac{1}{0}}\bigg) = \pi$. Similarly, when $q_{12}=\Ket{0}$, the probabilities of $q_{11}=\Ket{0}$ and $q_{11}=\Ket{1}$ is computed as $P(q_{11}=\Ket{0}|q_{12}=\Ket{0}) = \dfrac{P(q_{12}=\Ket{0},q_{11}=\Ket{0})}{P(q_{12}=\Ket{0})} = \dfrac{0.19}{0.82} = 0.232$ and $P(q_{11}=\Ket{1}|q_{12}=\Ket{0}) = 1-\dfrac{P(q_{12}=\Ket{0},q_{11}=\Ket{0})}{P(q_{12}=\Ket{0})} = 0.768$. Therefore, the rotation angle to show the conditional probability values when $q_{12}=\Ket{0}$ is $\theta_{q_{11}, q_{12}=\Ket{0}, q_{15}=\Ket{1}} = 2\times \tan^{-1}\bigg(\sqrt{\dfrac{0.768}{0.232}}\bigg) = 2.1365$. In this way, the two-qubit rotation gate $U$ is decomposed into a combination of single and two-qubit ($CR_Y$) gates. Since $U$ is implemented when $q_{15}=\Ket{1}$, the controlled-rotations in $U$ gate decomposition become controlled-controlled rotations with B as an additional control qubit. Following Section \ref{subsec:2node_child}, the controlled-controlled rotations are implemented using an ancilla qubit. Similar decomposition procedure is followed to implement the two-qubit $U$ gate when $q_{15}=\Ket{0}$. In this way, the conditional probabilities associated with CH are realized. This procedure is then repeated to show the conditional probability values of three-level nodes LM, M, R, CHN, and IFR. Thus, the quantum circuit of the 9-node naive Bayes classifier is constructed using the C-QBN approach.

\begin{table}[htb]
\begin{center}
\setlength{\arrayrulewidth}{0.2mm}
\setlength{\tabcolsep}{7pt}
\renewcommand{\arraystretch}{1}
\caption{Comparison of marginal probabilities in the 9-node naive Bayes classifier with Netica and C-QBN}
\label{tab:9node}
\scalebox{0.95}{
\begin{tabular}{|c|c|c|c|c|}
\hline
& \textbf{Netica} & \multicolumn{3}{c|}{\textbf{C-QBN}} \\
\hline
Value & Probability & Mean & SD & 95\% CI \\
\hline
B=0 & 0.500  & 0.502	&	0.005	&	[0.498, 0.505]\\
\hline
AU=0 &  0.595  &	0.599	&	0.005	&	[0.595, 0.603]\\
\hline
IT=0 &  0.565 &	0.564	&	0.005	&	[0.560, 0.567]\\
\hline
CH=0   & 0.240  &  	0.243	&	0.008	&	[0.237, 0.249]\\
\hline
CH=1 & 0.630 & 0.627	&	0.008	&	[0.621, 0.633]\\
\hline
LM=0  & 0.230  & 0.231	&	0.006	&	[0.226, 0.235]\\
\hline
LM=1 & 0.635 & 0.637	&	0.004	&	[0.634, 0.640]\\
\hline
M=0 & 0.295  &  0.296	&	0.004	&	[0.293, 0.299]\\
\hline
M=1 & 0.595 & 0.596	&	0.005	&	[0.592, 0.600]\\
\hline
R=0 & 0.395  & 	0.394	&	0.006	&	[0.390, 0.398]\\
\hline
R=1 & 0.475 & 0.476	&	0.005	&	[0.472, 0.480]\\
\hline
CHN=0 & 0.255   & 	0.258	&	0.005	& [0.254, 0.261]\\
\hline
CHN=1 & 0.585 & 0.584	&	0.006	&	[0.579, 0.588]\\
\hline
IFR=0 &  0.140 & 	0.138	&	0.004	&	[0.136, 0.141]\\
\hline
IFR=1 & 0.720 & 0.721	&	0.004	&	[0.718, 0.724]\\
\hline
\end{tabular}
}
\end{center}
\end{table}

\begin{figure}[H]
\centering
\includegraphics[width=0.8\columnwidth]{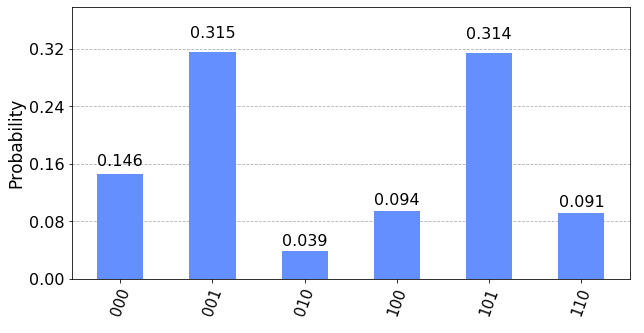}
\caption{Simulation results of nodes B and CH in the 9-node naive Bayes classifier using Qiskit with 8192 shots (quantum states on the horizontal axis and  joint probabilities on the vertical axis)} 
\label{fig: B-CHhistogram}
\end{figure}

\textbf{Circuit simulation and results:} Similar to the previous examples, we ran the circuit 10 times, each with 8192 shots. The mean, standard deviation, and the 95\% CI of the marginal probabilities are given in Table \ref{tab:9node}, from which it can be observed that the probability values from Netica fall within the 95\% CI obtained from the C-QBN approach. 
For variables with three states (CH, LM, M, R, CHN, IFR), we provided the probabilities of two states as the probability of the third state can be computed using the probabilities of the given states. For example, $P(M=2)=1-P(M=0)-P(M=1)$. These results help conclude that the proposed circuit construction approach can represent the 9-node Naive Bayes classifier. The histogram in Figure \ref{fig: B-CHhistogram} shows different quantum state probabilities in the partial circuit using nodes B and CH. 

Each state represents a joint probability of the corresponding variables. The first value in each of the states corresponds to B and the other two are used to represent the three states of CH. Since CH has only three states irrespective of the value of $B$, the probability of state $\Ket{11}$ is fixed at zero. Therefore, the probabilities associated with $\Ket{011}$ and $\Ket{111}$ states are equal to zero and do not appear in the histogram.

We ran each of the three examples to compute the simulation time, and we observed that the simulation time is not constant but changes across each run. The ranges of the simulation times for analyzing the 4-node, 10-node and 9-node BNs are [15ms, 23ms], [28ms, 32ms] and [50ms, 66ms] respectively. Here, ``ms" stands for milliseconds. Since Netica is a commercial software, we were unable to extract the exact computation time. One of the main advantages of the quantum Bayesian networks is that there have been proven computational benefits in  forward and inverse analysis through amplitude amplification and estimation \mbox{\citep{low2014quantum,woerner2019quantum}}.

\textbf{Noise analysis:}
In order to understand the effect of hardware noise on the marginal probabilities, we ran the Bayesian network examples after incorporating the hardware noise from real quantum hardware (from IBM) and compared the results when run without incorporating any hardware noise. We considered noise models from seven publicly available quantum hardware: Burlington, Vigo, Ourense, London, Essex, Yorktown and Melbourne \mbox{\citep{IBMQE}}. All the hardware except for Melbourne have five qubits whereas Melbourne has fifteen qubits. Moreover, in these hardware, there exists limited connectivity between qubits, i.e., every qubit is not connected to every other qubit \mbox{\citep{IBMQE}}. The connectivity is important when implementing the two-qubit gate (CNOT). CNOT gate can not be applied between qubits that are not connected to each other in the hardware. Therefore, in addition to noise models, we incorporated the connectivity information to simulate the execution on real hardware.

Since the first example requires five qubits (four qubits for the four nodes in the Bayesian network and one ancilla qubit),  it can be run on all the hardware whereas the second example which requires eleven qubits (including one ancilla qubit) can only be run on Melbourne as the remaining hardware have only five qubits. The third example requires sixteen qubits and unfortunately as of July 2020, we do not have public access to a hardware which can handle sixteen qubits.

The noise models get updated whenever the quantum hardware are calibrated \mbox{\citep{IBMQE}}.The noise models that we used for analysis correspond to the noise models on July 10, 2020.

Table \mbox{\ref{tab:4bnnoise}} provides the marginal probabilities of the four nodes obtained from Netica, with and without incorporating any hardware noise in Qiskit. Since there exists variation across multiple runs, we provide the mean and standard deviation values for analysis on the simulator. The standard deviation values are available in the parenthesis. The difference between the results obtained from Qiskit and Netica are quantified using the the Root Mean Square Percentage Error (RMSPE), using Eq.\mbox{\ref{eq: epsilon}}.

\begin{equation}
\label{eq: epsilon}
\epsilon_T = 100\%\sqrt{\dfrac{1}{n} \sum_{i}  \bigg(\dfrac{p_i^t - \bar{p_i}}{p_i^t}\bigg)^2}
\end{equation}

where $\epsilon_T$ is the RMSPE, $p_i^t$ and $\bar{p_i}$ are the true and expectation values (over 10 runs), and $n$ represents the number of nodes in the Bayesian network. The true values are obtained from classical analysis using Netica software.

\begin{figure}[htbp]
    \centering
    \includegraphics[width=0.99\columnwidth]{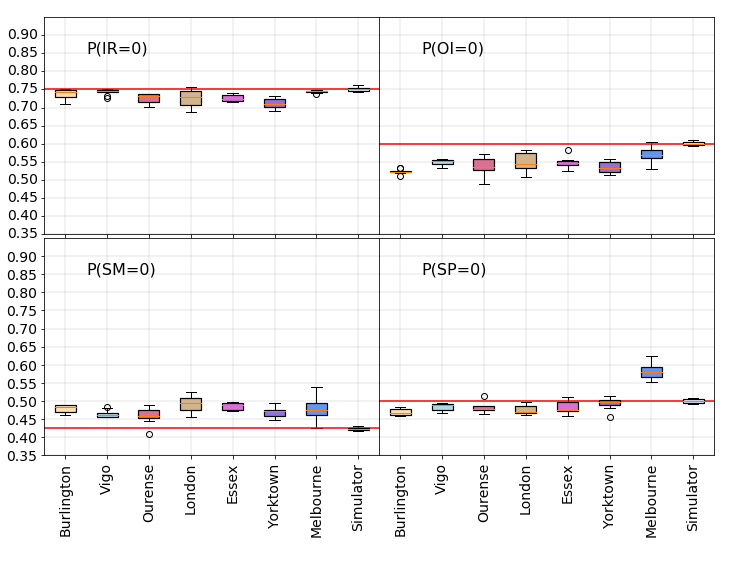}
    \caption{Box plots associated with the marginal probabilities of the 4-node Bayesian network on Qiskit with noise models from different IBM QX hardware, without any noise models, and marginal probabilities from Netica (red lines). Each of the four subplots corresponds to a node in the Bayesian network (IR, OI, SM, SP). Various hardware models are available on the horizontal axis (Simulator represents the case without any noise model) and probabilities on the vertical axis.}
    \label{fig:4nodebox}
\end{figure}

It can be observed that the simulator results without any hardware noise are close to those of the true values (RMSPE of 0.11\%). However, the results were noticeably different when the hardware noise models are incorporated with RMSPE ranging from 6.44\% (Vigo) to 10.8\% (Melbourne). The mean and the standard deviation values, box plots showing the variation in marginal probabilities with various hardware noise models and without noise models along with the true values are provided in Figure \ref{fig:4nodebox}.

\begin{table}[htbp]
\centering
\caption {Mean and standard deviation values of marginal probabilities over 10 runs of the 4-node Bayesian network on Qiskit with noise models from different IBM QX hardware, without any noise models, and marginal probabilities from Netica}
\scalebox{0.9}{
  \begin{tabular}{|p{1.5cm}|p{1.7cm}|p{1.7cm}|p{1.7cm}|p{1.7cm}|p{1.3cm}|}
\hline

      & \multicolumn{1}{p{5.5em}|}{\textbf{P(IR=0)}} &  \multicolumn{1}{p{5.5em}|}{\textbf{P(SM=0)}} & \multicolumn{1}{p{5.5em}|}{\textbf{P(OI=0)}} & \multicolumn{1}{p{5.5em}|}{\textbf{P(SP=0)}} &
      \multicolumn{1}{p{4em}|}{\textbf{RMSPE}} \\

    \hline
    Netica & 0.750 &0.425 & 0.600 & 0.499 & --- \\
    \hline
    Simulator & 0.750(0.008)  & 0.425(0.006) & 0.601(0.003) & 0.499(0.004) & 0.11\% \\
    \hline
    \multicolumn{6}{|c|}{Noise Models} \\
    \hline
    Burlington & 0.738(0.013) & 0.479(0.010) & 0.524(0.007)  & 0.469(0.010) & 9.50\% \\
    Vigo       & 0.743(0.007) & 0.464(0.010) & 0.550(0.008)  & 0.484(0.009) & 6.44\% \\
    Ourense    & 0.726(0.014) & 0.459(0.021) & 0.538(0.024)  & 0.482(0.013) & 6.98\% \\
    London     & 0.726(0.024) & 0.491(0.021) & 0.550(0.024)  & 0.476(0.012) & 9.29\% \\
    Essex      & 0.726(0.009) & 0.487(0.009) & 0.548(0.015)  & 0.483(0.016) & 8.77\% \\
    Yorktown   & 0.711(0.014) & 0.469(0.013) & 0.535(0.015)  & 0.493(0.015) & 7.97\% \\
    Melbourne  & 0.744(0.003) & 0.480(0.031) & 0.568(0.022)  & 0.581(0.020) & 10.80\% \\
    \hline
    \end{tabular}%
}
  \label{tab:4bnnoise}%
\end{table}%

\begin{table}[htbp]
\centering
\caption{Mean and standard deviation values of marginal probabilities over 10 runs of the 10-node Bayesian network on the simulator with and without including Melbourne noise model, and marginal probabilities from Netica}
\scalebox{0.9}{
    \begin{tabular}{|l|l|l|l|}
    \hline
\textbf{Marginal} & \multicolumn{1}{p{4em}|}{\textbf{Netica}} & \multicolumn{1}{p{6em}|}{\textbf{Simulator}} & \multicolumn{1}{p{6em}|}{\textbf{Melbourne}}   \\
    \hline
      $P(X_1=0)$ & 0.431 & 0.432 (0.007) & 0.501 (0.042) \\
    \hline
      $P(X_2=0)$ & 0.863 & 0.864 (0.003) & 0.431 (0.031) \\
    \hline
      $P(X_3=0)$ & 0.976 & 0.976 (0.002) & 0.749 (0.048) \\
    \hline
      $P(X_4=0)$ & 0.570 & 0.570 (0.006) & 0.410 (0.044) \\
    \hline
      $P(X_5=0)$ & 0.527 & 0.525 (0.004) & 0.577 (0.053) \\
    \hline
      $P(X_6=0)$ & 0.980 & 0.980 (0.001) & 0.895 (0.059) \\
    \hline
      $P(X_7=0)$ & 0.977 & 0.976 (0.001) & 0.867 (0.046) \\
    \hline
      $P(X_8=0)$ & 0.026 & 0.027 (0.001) & 0.492 (0.047) \\
    \hline
      $P(X_9=0)$ & 0.956 & 0.956 (0.002) & 0.590 (0.052) \\
    \hline
      $P(X_{10}$=0) & 0.240 & 0.239 (0.006) & 0.559 (0.031) \\
    \hline
    RMSPE & ---  &   0.6\%    & 557.2\%   \\
    \hline
    \end{tabular}
}
\label{tab:10bnnoise}%
\end{table}%

\begin{figure}[htbp]
    \centering
    \includegraphics[width=0.49\columnwidth]{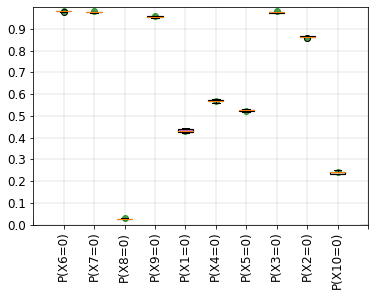}
    \includegraphics[width=0.49\columnwidth]{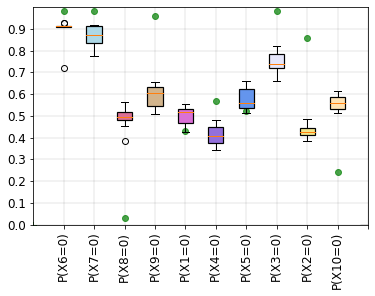}
    \caption{Box plots associated with marginal probabilities of the 10-node Bayesian network on Qiskit without adding any noise model (left) and with noise model from Melbourne (right) compared with the results from Netica(green points). The variables are on the horizontal axis and their probabilities are on the vertical axis.
    } 
    \label{fig:box10bn}
\end{figure}

Similar to the 4-node example, Table \ref{tab:10bnnoise} provides the mean and standard deviation values of the the marginal probabilities of various nodes from Netica, with and without incorporating any hardware noise in Qiskit. As metioned above, we provide results with Melbourne noise model as the 10-node Bayesian network cannot be run with other available hardware. It can be observed that adding the noise model produces highly erroneous results with 557.2\% RMSPE when compared to 0.6\% RMSPE without any noise model. Figure \ref{fig:box10bn} provides the box plots of the marginal probabilities from simulator without any hardware noise and with Melbourne noise model along with the true values obtained from Netica. Figure \ref{fig:box10bn} shows the wide variation between the results from Netica and those obtained with Melbourne noise model. From this example, we can infer that even though the proposed approach can be used to represent large Bayesian networks, their implementation on real hardware can produce erroneous results due to hardware noise.\\

\section{Conclusion}
\label{sec:conclusion}

This paper detailed the design of a quantum circuit to represent a generic discrete Bayesian network with nodes that may have two or more states. The quantum circuit design follows three steps. The first step is to map a Bayesian network node to one or more qubits depending on the number of states. The second step is mapping the marginal or conditional probabilities of nodes to probability amplitudes/probabilities associated with the qubits to be in $\Ket{0}$ and $\Ket{1}$ states. The third step is to realize the required probability amplitudes using single-qubit and (multi-qubit) controlled rotation gates. We used ancilla qubits for the implementation of multi-qubit rotation gates. When a node is mapped to more than one qubit, the multi-qubit rotations required to realize the required probabilities are decomposed into a combination of single-qubit and multi-qubit controlled rotations. 

The proposed approach was demonstrated with three Bayesian networks: a Bayesian network with four nodes and each with two states used for an oil company stock prediction, a Bayesian network with ten nodes and each with two states used for liquidity risk assessment, and a Naive Bayes classifier with nine nodes (eight features). Of the nine nodes, three nodes had two states and six nodes had three states. The quantum circuits are designed and simulated on Qiskit \citep{mckay2018qiskit}, which is a Python-based simulator for quantum computing. We simulated each circuit with 8192 shots, and calculated the marginal probabilities associated with each node. Since the results from quantum circuit are stochastic, we repeated the simulations 10 times, each time with 8192 shots. Using the results from 10 simulations, we estimated the 95\% confidence intervals. To validate the simulation results, we simulated the Bayesian networks using a classical Bayesian network software (Netica) and tested if the Qiskit results match the results from the classical software. We found that the marginal probabilities of all the nodes obtained from the classical implementations were within the 95\% confidence intervals obtained from Qiskit. 

All the quantum circuits in this work were designed manually. Future work should consider automating the design of quantum circuit for a given Bayesian network. We used a simulation platform in this work to validate the accuracy of the methods. In future, we will implement the proposed methods on real quantum computers, and study the effect of hardware noise on the circuit implementation.  
\bibliographystyle{model5-names}\biboptions{authoryear}
\bibliography{references}


\begin{figure}[htb]
    \centering
    \includegraphics[scale=0.225]{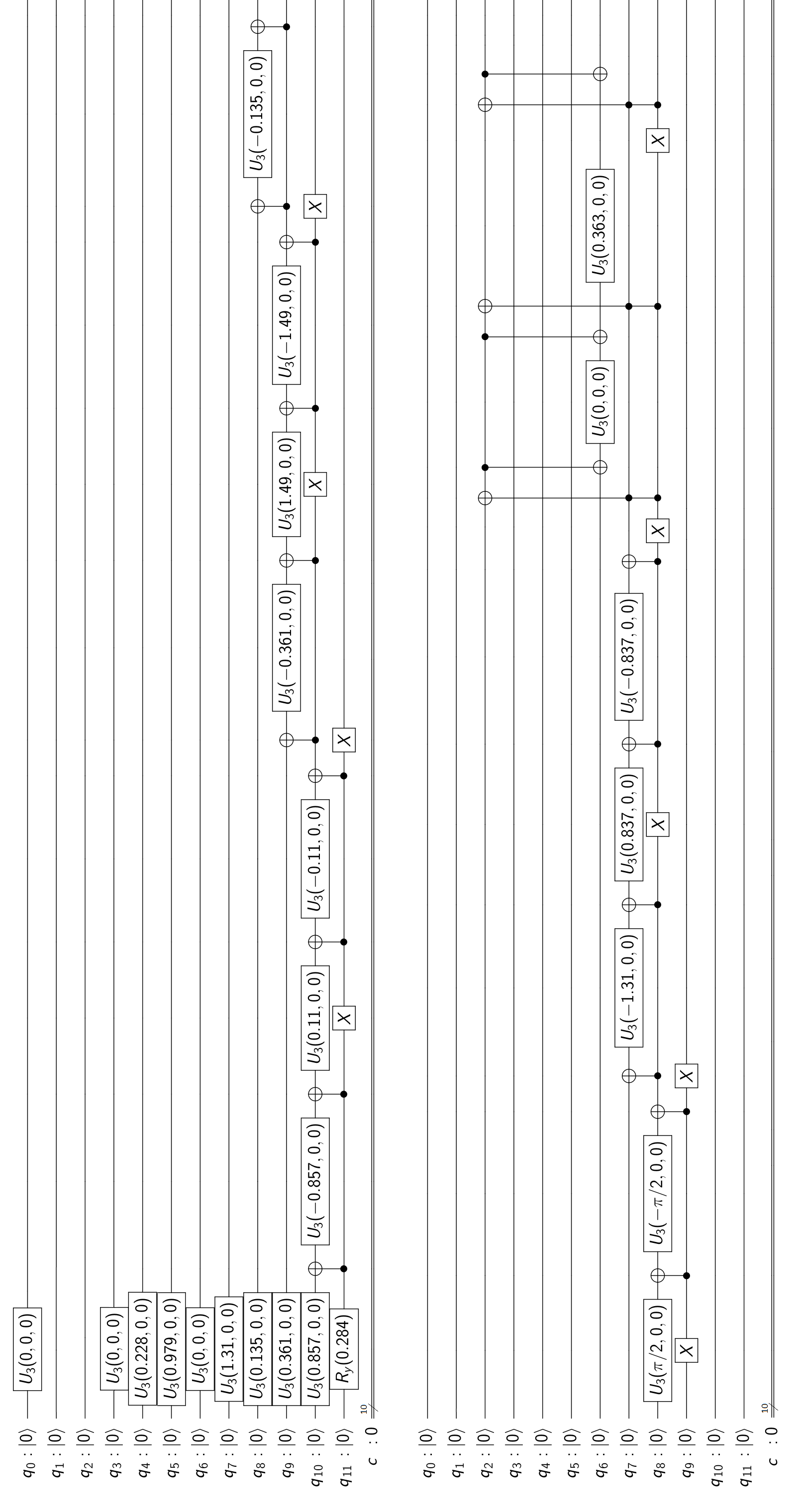}
    \caption{Quantum circuit of the 10-node liquidity risk assessment BN. Variables $X_6$, $X_7$, $X_8$, $X_9$, $X_1$, $X_4$, $X_5$, $X_3$, $X_2$ and $X_{10}$ are mapped to $q_{11}$, $q_{10}$, $q_{9}$, $q_{8}$, $q_{7}$, $q_{6}$, $q_{5}$, $q_{4}$, $q_{3}$, and $q_{0}$ respectively, $q_1$ and $q_0$ are the ancilla qubits, and $c$ represents classical bits.} 
    \label{fig:10nodeBN_cqbn}
\end{figure}

\begin{figure}[htb]
\ContinuedFloat
    \centering
    \includegraphics[scale=0.225]{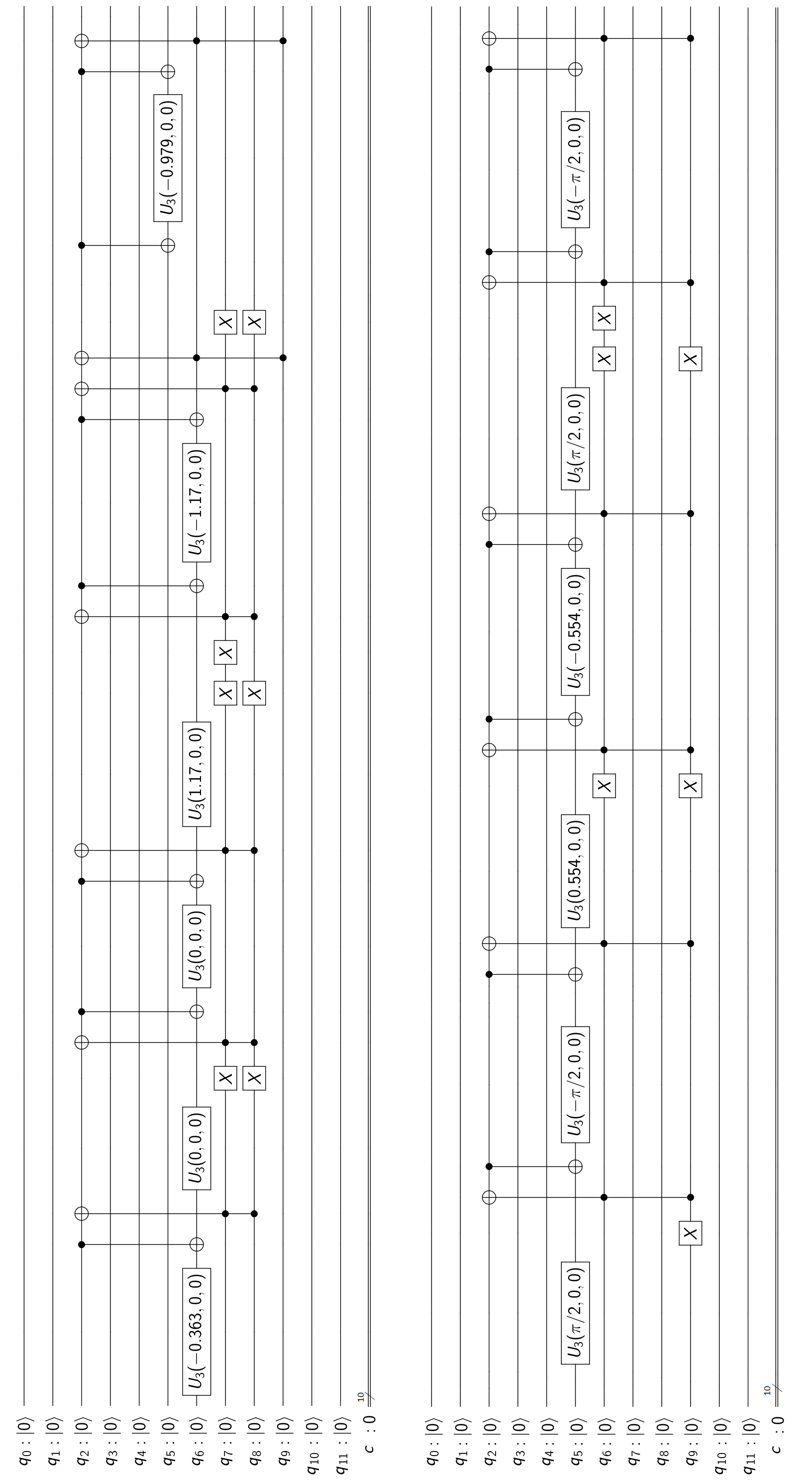}
    \caption{Quantum circuit of the 10-node liquidity risk assessment BN. Variables $X_6$, $X_7$, $X_8$, $X_9$, $X_1$, $X_4$, $X_5$, $X_3$, $X_2$ and $X_{10}$ are mapped to $q_{11}$, $q_{10}$, $q_{9}$, $q_{8}$, $q_{7}$, $q_{6}$, $q_{5}$, $q_{4}$, $q_{3}$, and $q_{0}$ respectively,  $q_1$ and $q_0$ are the ancilla qubits, and $c$ represents  classical bits. (contd)}
    \vspace{5cm}
\end{figure}

\begin{figure}[htb]
\ContinuedFloat
    \centering
    \includegraphics[scale=0.225]{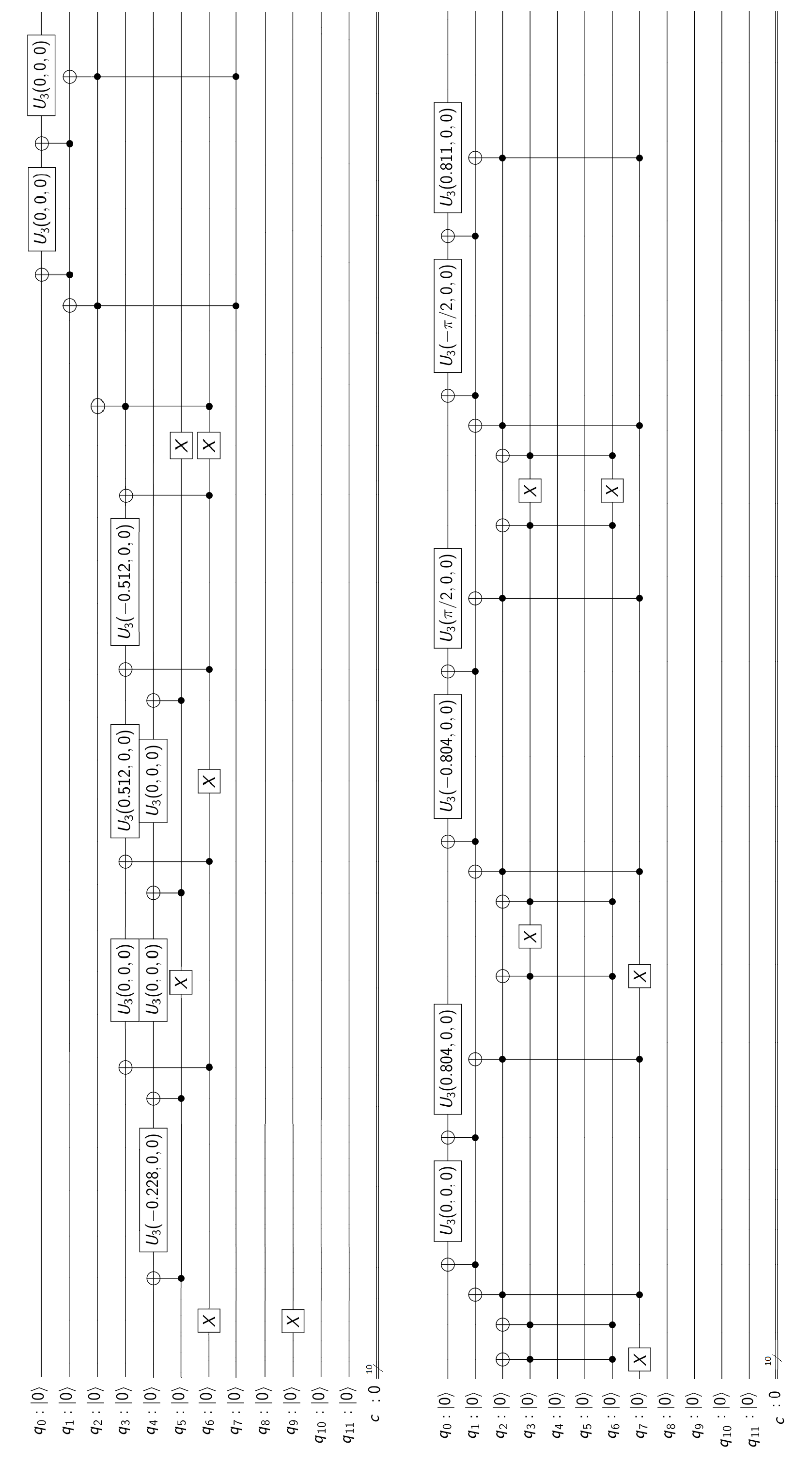}
    \caption{Quantum circuit of the 10-node liquidity risk assessment BN. Variables $X_6$, $X_7$, $X_8$, $X_9$, $X_1$, $X_4$, $X_5$, $X_3$, $X_2$ and $X_{10}$ are mapped to $q_{11}$, $q_{10}$, $q_{9}$, $q_{8}$, $q_{7}$, $q_{6}$, $q_{5}$, $q_{4}$, $q_{3}$, and $q_{0}$ respectively, $q_1$ and $q_0$ are the ancilla qubits, and $c$ represents  classical bits. (contd)}
    \vspace{5cm}
\end{figure}

\begin{figure}[htb]
\ContinuedFloat
    \centering
    \includegraphics[scale=0.225]{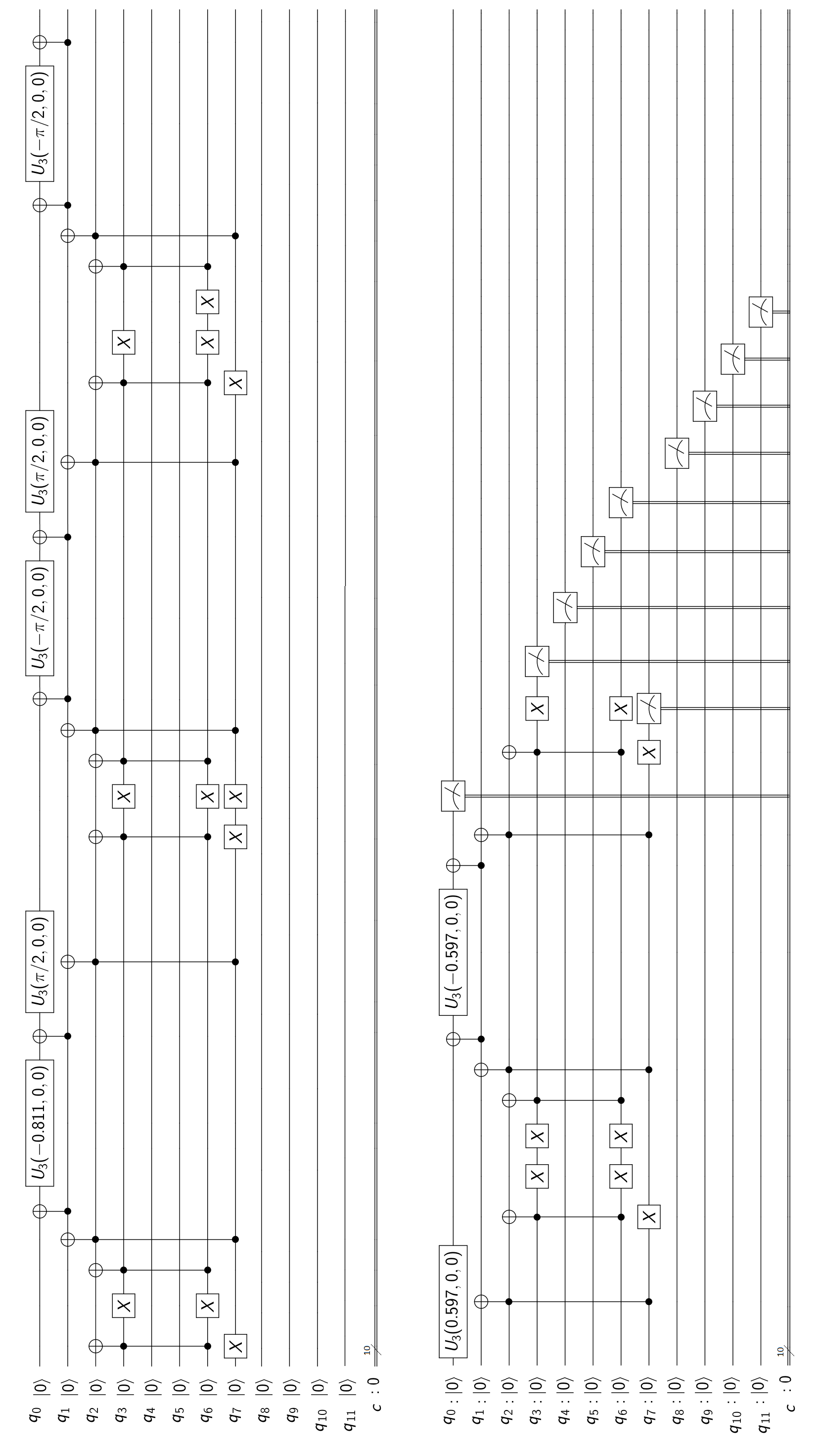}
    \caption{Quantum circuit of the 10-node liquidity risk assessment BN. Variables $X_6$, $X_7$, $X_8$, $X_9$, $X_1$, $X_4$, $X_5$, $X_3$, $X_2$ and $X_{10}$ are mapped to $q_{11}$, $q_{10}$, $q_{9}$, $q_{8}$, $q_{7}$, $q_{6}$, $q_{5}$, $q_{4}$, $q_{3}$, and $q_{0}$ respectively, $q_1$ and $q_0$ are the ancilla qubits, and $c$ represents  classical bits. (contd)}
    \vspace{5cm}
\end{figure}

\newpage

\end{document}